\documentclass[useAMS,usenatbib]{mn2e}
\usepackage{graphicx}%for EPS, PS files
\usepackage{amssymb}
\usepackage{enumerate}

\footnotesize
\newdimen\digitwidth    %define ! a one digit width for tables
\setbox0=\hbox{\rm0}
\digitwidth=\wd0
\catcode`!=\active
\def!{\kern\digitwidth}
\normalsize

\newcommand{\psr}{PSR~J1614$-$2230}

\title[On the accretion, spin-up and ages of MSPs]
      {Formation of millisecond pulsars with {CO} white dwarf companions
      -- II. Accretion, spin-up, true ages and comparison to MSPs with {He} white dwarf companions}

\author[Tauris, Langer \& Kramer]{
  T.~M.~Tauris,$^1$$^,$$^2$ N.~Langer$^1$ and M.~Kramer$^2$$^,$$^3$
  \\
  $^{1}$Argelander-Insitut f\"ur Astronomie, Universit\"at Bonn, Auf
  dem H\"ugel 71, 53121 Bonn, Germany 
  \\
  $^{2}$Max-Planck-Institut f\"ur Radioastronomie, Auf dem H\"ugel 69,
  53121 Bonn, Germany
  \\
  $^{3}$Jodrell Bank Centre for Astrophysics, University of Manchester,
  Oxford Rd, Manchester M13 9PL, UK 
  \\
  }
\date{Accepted 2012 June 3}
\begin{document}

\maketitle
%\newcommand{\setthebls}{
%                 de-comment this line for double spacing:
%\baselineskip=20pt
%}
%\setthebls

\begin{abstract} 
Millisecond pulsars are mainly characterised by their spin periods, B-fields and 
masses -- quantities which are largely affected by previous interactions with a companion star in a binary system.
In this paper, we investigate the formation mechanism of millisecond pulsars 
by considering the pulsar recycling process in both intermediate-mass {X}-ray binaries (IMXBs)
and low-mass {X}-ray binaries (LMXBs). The IMXBs mainly lead to the formation of binary
millisecond pulsars with a massive carbon-oxygen ({CO}) or an oxygen-neon-magnesium white dwarf ({ONeMg}~WD) companion,
whereas the LMXBs form recycled pulsars with a helium white dwarf ({He}~WD) companion.
We discuss the accretion physics leading to the spin-up line in the $P\dot{P}$--diagram 
and demonstrate that such a line cannot be uniquely defined.
We derive a simple expression for the amount of accreted mass needed for any given pulsar 
to achieve its equilibrium spin and apply this to explain the observed differences of the spin distributions
of recycled pulsars with different types of companions.
From numerical calculations we present further evidence for significant loss of rotational energy in accreting {X}-ray millisecond pulsars 
in LMXBs during the Roche-lobe decoupling phase (Tauris~2012) 
and demonstrate that the same effect is negligible in IMXBs.
We examine the recycling of pulsars with {CO}~WD companions via Case~BB 
Roche-lobe overflow (RLO) of naked helium stars in post common envelope binaries.
We find that such pulsars typically accrete of the order $0.002-0.007\,M_{\odot}$ 
which is just about sufficient to explain their observed spin periods.
We introduce isochrones of radio millisecond pulsars in the $P\dot{P}$--diagram to follow their spin evolution 
and discuss their true ages from comparison with observations.
Finally, we apply our results of the spin-up process to complement our investigation of the massive pulsar \psr\ from
Paper~I, confirming that this system formed via stable Case~A RLO in an IMXB and
enabling us to put new constraints on the birth masses of a number of recycled pulsars.  
\end{abstract}


%%%%%%%%%%%%%%%%%%%%%%%%%%%%%%%%%%%%%%%%%%%%%%%%%%%%%%%%%%%%%%%%%%%%%%%%%%%%%%%%

\section{Introduction}
\label{sec:Intro}

Binary millisecond pulsars (BMSPs) represent the advanced phase of stellar evolution in close, interacting binaries. Their observed orbital
and stellar properties are fossil records of their evolutionary history. Thus one
can use binary pulsar systems as key probes of stellar astrophysics.
It is well established that the neutron star component in binary millisecond pulsar systems forms first, descending from the
initially more massive of the two binary zero-age main sequence (ZAMS) stars. 
The neutron star is subsequently spun up to a high spin frequency
via accretion of mass and angular momentum once the secondary star evolves
\citep{acrs82,rs82,bv91}. In this recycling phase the
system is observable as a low-mass {X}-ray binary \citep[e.g.][]{hay85,nag89,bcc+97} and towards the end of this phase
as an {X}-ray millisecond pulsar \citep{wv98,asr+09,pw12}.
Although this standard formation scenario is commonly accepted
many aspects of the mass-transfer process and the accretion physics are
still not understood in detail \citep{lv06}. Examples of such ambiguities include the accretion disk structure, 
the disk-magnetosphere transition zone, the accretion efficiency, 
the decay of the surface B-field of the neutron star 
and the outcome of common envelope evolution.  

The current literature on the recycling process of pulsars is based on a somewhat simplified treatment
of both the accretion process and the so-called equilibrium spin theory, as well as the application of
a pure vacuum magnetic dipole model for estimating the surface B-field strength of a radio pulsar.
These simplifications become a problem when trying to probe the formation and the evolution of observed
recycled radio pulsars located near the classical spin-up line for Eddington accretion in the $P\dot{P}$--diagram 
\citep[e.g.][]{faa+11}. 
In this paper we discuss the concept and the location of the spin-up line and investigate to which
extent it depends on the assumed details of the accretion disk/magnetosphere interactions and the magnetic inclination angle
of the pulsar. Furthermore, we include the plasma filled magnetosphere description by \citet{spi06} 
for determining the surface B-field strengths of pulsars 
in application to spin-up theory and compare to the case of using the standard vacuum dipole model.

A key quantity to understand the formation of any given recycled radio pulsar is its spin period  
which is strongly related to the amount of mass accreted. The amount of accumulated mass is again determined
by the mass-transfer timescale of the progenitor {X}-ray binary (and partly by the mode of the mass transfer and mass loss from the system),
and hence it depends crucially on the nature of the donor star. 
At present the scenario is qualitatively well understood where low-mass {X}-ray binaries (LMXBs)
and intermediate-mass {X}-ray binaries (IMXBs) in general lead to the formation of pulsars with 
a helium white dwarf ({He}~WD) or a carbon-oxygen white dwarf ({CO}~WD) companion, respectively \citep[e.g.][]{tvs00,prp02,del08,tau11}. 
However, here we aim to quantify this picture in more detail by re-analysing the spin-up process.
As an example, we investigate in this paper the possibilities of spinning up a neutron star 
from Case~BB RLO leading to a mildly recycled pulsar with a {CO}~WD companion. 
We also discuss the increasing number of systems which apparently do not fit the characteristics of these 
two main populations of binary pulsars. 

We follow the pulsar spin evolution during the final phase of the mass transfer by including
calculations of the torque acting on an accreting pulsar during the
Roche-lobe decoupling phase \citep[RLDP;][]{tau12} for both an LMXB and an IMXB and compare the results. 
To complete the full history of pulsar spin evolution we subsequently  
consider the true ages of recycled radio pulsars (which are intimately related to their spin evolution) 
by calculating isochrones and discussing the distribution of recycled pulsars in the $P\dot{P}$--diagram.  

Although accreting {X}-ray millisecond pulsars (AXMSPs) are believed to be progenitors of BMSPs, 
all of the 14 observed AXMSPs have orbital periods less than one day whereas fully recycled radio
BMSPs are observed with orbital periods all the way up to a few hundred days \citep[see also][]{tau12}.
This puzzle will be investigated in a future paper. For the evolution of ultra-compact {X}-ray binaries (UCXBs)
leading to AXMSPs, and later to the formation of tight-orbit recycled radio BMSPs with ultra-light 
($<0.08\,M_{\odot}$) companions, we refer to
\citet{prp02,db03,vvp05,vnv+12}. In this paper we focus on the formation and evolution of BMSPs with
regular {He}~WDs and {CO}/{ONeMg}~WDs.

The discovery of \psr\ \citep{hrr+05,dpr+10} plays an important role for understanding BMSP formation. 
This pulsar system is interesting since it has a unique combination of a short pulsar spin period
of 3.2~ms and a massive {CO}~WD companion. A rapidly spinning pulsar is usually associated with
a long phase of mass-transfer evolution in an LMXB, whereas  
a {CO}~WD companion in a relatively close orbit (8.7~days) is evidently the outcome of an IMXB evolution.
A possibly solution to this paradox is that \psr\ evolved from Case~A RLO of an IMXB which results in both a relatively long-lived
{X}-ray phase ($>10^7\,{\rm yr}$), needed to spin~up the pulsar effectively, and leaving behind a {CO}~WD. 
In this case \psr\ is the first BMSP known to have evolved via this path. Indeed, in     
\citet{tlk11}, hereafter Paper~I, we investigated the progenitor evolution of \psr\  
with emphasis on the {X}-ray phase where the binary contained a neutron star and a donor star.
We found two viable possibilities for the formation of
the \psr\ system: either it contained a $2.2-2.6\,M_{\odot}$ asymptotic giant branch donor star and evolved through 
a common envelope and spiral-in phase initiated by Case~C RLO,
or it descended from a close binary system with 
a $4.0-5.0\,M_{\odot}$ main sequence donor star via Case~A RLO as hinted above. 
The latter scenario was also found by \citet{lrp+11}.
The fact that \psr\ was spun-up to (less than) 3.2~ms could indeed hint which one of the two formation scenarios is most likely.
In order to test this idea and to further distinguish between Case~A and Case~C
we turn our attention, here in Paper~II, to the spin dynamics of this pulsar in the
recycling process.  

As discussed in Paper~I, the distribution of neutron star birth masses is an important probe of both stellar evolution,  
binary interactions and explosion physics. 
For a number of BMSPs we are now able to put constraints on the birth mass of the pulsar given the derived 
amount of mass needed to spin up the observed recycled pulsar. In particular, it is of interest to see 
if we can identify further pulsars showing evidence of being born massive ($\sim\!1.7\,M_{\odot}$)
like \psr.

In order to understand the many different observational properties of BMSPs we have combined a variety of subtopics here with the
aim of presenting a full picture of the subject.
Given the many facets included, our new findings and the resulting length of the manuscript, 
we have chosen throughout this paper to finalize individual subtopics with a direct comparison to observational data
followed by discussions in view of our theoretical modelling. 

Our paper is structured in the follow way: We begin with a brief, updated review of the formation channels of
BMSPs with {CO}~WDs (Section~\ref{sec:formation}) and present a 
summary of the latest observational data in Section~\ref{sec:observations}. In this section we also demonstrate an emerging
unified picture of pulsar formation history which, however, is challenged by a number of interesting systems
which share expected properties of both an LMXB and an IMXB origin.
In Section~\ref{sec:spinup} we investigate the recycling process in general 
with a focus on the location of the spin-up line in the $P\dot{P}$--diagram 
and also relate the initial spin of a rejuvenated pulsar to the amount of mass accreted.
The theoretical modelling is continued in Section~\ref{sec:RLDP} where we highlight the effects
of the Roche-lobe decoupling phase on the spin evolution of recycled pulsars.
In Section~\ref{sec:obs-spin} our results are compared to the observed spin period distributions and 
in Section~\ref{sec:CEspin} we investigate if BMSPs with {CO}~WD companions 
obtained their fast spin periods {\it after} a common envelope evolution. 
In Section~\ref{sec:trueages} we discuss our results in a broader context in relation to 
the spin evolution and the true ages of millisecond radio pulsars. 
In Section~\ref{sec:1614-2230} we continue our discussion from Paper~I on the formation and evolution of \psr\
and in Section~\ref{sec:NSmass} we return to the question of neutron star birth masses.
Our conclusions are summarized in Section~\ref{sec:summary}.
Finally, in the Appendix we present a new tool to identify the most likely nature
of the companion star to any observed binary pulsar. 

%%%%%%%%%%%%%%%%%%%%%%%%%%%%%%%%%%%%%%%%%%%%%%%%%%%%%%%%%%%%%%%%%%%%%%%%%%%%%%%%

\section{Formation of BMSPs with {CO}~WD companions}\label{sec:formation}
According to stellar evolution theory one should expect radio pulsars to exist in binaries with a variety of different companions: 
white dwarfs ({He}~WDs, {CO}~WDs and {ONeMg}~WDs), neutron stars, black holes, ultra~light semi-degenerate dwarfs (i.e. substellar companions
or even planets),
helium stars, main sequence stars and, for very wide systems, sub-giant and giant stars \citep[][and references therein]{bv91,tv06}.
Of these possibilities, pulsars orbiting black holes, helium stars and giant stars still remain to be detected.\\
The majority of radio BMSPs have {He}~WD companions. The formation of these systems is mainly channeled through LMXBs
and have been well investigated in previous studies \citep[e.g.][]{wrs83,ps88,ps89,rpj+95,esa98,ts99,prp02,ndm04,vvp05,del08}. 
These systems have orbital periods between
less than 0.2~days and up to several hundred days. One of the most striking features of these systems is the relation between white dwarf
masses and orbital periods \citep[e.g.][]{ts99}.
However, BMSPs with {He}~WD companions may also form in IMXBs if Case~A RLO is not initiated too late during the main sequence evolution
of the donor star \citep{tvs00,prp02}. In Section~\ref{sec:observations} we identify 6~BMSPs which may have formed via this channel.

BMSP systems with relatively heavy WDs ({CO} or {ONeMg} WDs) all have observed orbital periods
$\le 40\,{\rm days}$. For these systems there are a number of suggested formation channels which are briefly summarized below 
-- see Paper~I, and references therein, for a
more rigorous discussion. In order to leave behind a {CO}~WD, its progenitor must be more massive than
$3.0\,M_{\odot}$ if Roche-lobe overflow (RLO) is initiated while the donor star is still on the main sequence. If the
donor star is an asymptotic giant branch (AGB) star when initiating mass transfer it can, for example, have
a mass as low as $2.2\,M_{\odot}$ and still leave behind a $0.5\,M_{\odot}$ {CO}~WD remnant -- see fig.~1 in Paper~I.
{CO}~WDs in close binaries have masses between $0.33-1.0\,M_{\odot}$, whereas the {ONeMg}~WDs are slightly heavier
with masses of $1.1-1.3\,M_{\odot}$. The upper limit for the initial mass of donor stars in {X}-ray binaries leaving an {ONeMg}~WD is 
often assumed to be about 
$7-8\,M_{\odot}$ \citep{plp+04}, depending on the orbital period, metallicity, treatment of convection and the amount of convective overshooting. 
However, \citet{wl99} demonstrated that even stars with an initial mass in excess of $13\,M_{\odot}$ may form an {ONeMg}~WD 
if these progenitor stars are in a tight binary and thus lose their envelope at an early stage via Case~A RLO. 
Stars exceeding the upper threshold mass for producing an {ONeMg}~WD will leave behind a neutron star remnant via 
an electron capture supernova \citep{nom84,plp+04} 
or via a core-collapse supernova of type~Ib/c for slightly more massive stars. For single stars, or very wide binaries, the critical ZAMS mass for neutron star
production via type~II supernovae is about $10\,M_{\odot}$ \citep{zwh08}.\\
In this paper we consider all systems with either
a {CO} or an {ONeMg}~WD as {\it one} population, denoted by {CO}~WDs unless specified otherwise, and 
all scenarios listed below apply to BMSPs with both types of massive white dwarf companions.
However, the required progenitor star masses are, in general, somewhat larger for the {ONeMg}~WDs compared to the {CO}~WDs.
Despite these minor differences, the bottom line is that in all cases 
IMXBs are the progenitor systems of BMSPs with massive white dwarf companions. 

\subsection{IMXBs with Case~A RLO}
BMSPs with {CO}~WDs, which formed via Case~A RLO in IMXBs,
evolve from donor stars with masses $3-5\,M_{\odot}$ and orbital periods of a few days (\citet{tvs00,prp02}; Paper~I). The final orbital periods 
of the BMSPs are $5-20\,{\rm days}$. The mass-transfer phase consists of three parts: A1, A2 and AB. Phase~A1 is thermal 
time-scale mass transfer and lasts for about $1\,{\rm Myr}$. The majority of the donor envelope is transfered at a high
rate exceeding $10^{-5}\,M_{\odot}\,{\rm yr}^{-1}$. However, the accretion onto the neutron star is limited by the
Eddington luminosity, corresponding to an accretion rate of a few $10^{-8}\,M_{\odot}\,{\rm yr}^{-1}$, and thus the far
majority of the transfered material (up to 99.9\%) is ejected from the system. 
Phase A2 is driven by the nuclear burning of the residual hydrogen in the core. 
The resulting mass-transfer rate is very low ($<10^{-9}\,M_{\odot}\,{\rm yr}^{-1}$) and only a few $10^{-2}\,M_{\odot}$ of material
is transfered towards the accreting neutron star in a time-interval of 
$20-50\,{\rm Myr}$. Phase~AB is caused by expansion of the donor star while undergoing hydrogen shell burning. The mass-transfer rate is
$10^{-8}-10^{-7}\,M_{\odot}\,{\rm yr}^{-1}$ and lasts for $\sim\!10\,{\rm Myr}$. A few $0.1\,M_{\odot}$ is transfered towards the
accreting neutron star -- significantly more than in the previous two phases, A1 and A2.
As we shall see later on, this phase is responsible for spinning up the neutron star to become a millisecond pulsar.

\subsection{IMXBs with early Case~B RLO}
Wider orbit IMXB systems evolve from donor stars which are in the Hertzsprung~gap at the onset of the RLO.
The orbital periods at the onset of the Case~B RLO are about $3-10\,{\rm days}$ and donor masses 
of $2.5-5.0\,M_{\odot}$ are needed to yield a {CO}~WD remnant in a BMSP system \citep{tvs00,prp02}.
The final orbital periods of the BMSPs are about
$3-50\,{\rm days}$. The mass-transfer rate in the IMXB phase
is highly super-Eddington ($10^{-5}\,M_{\odot}\,{\rm yr}^{-1}$) and thus the far majority
of the transfered material is ejected. However, a few $10^{-2}\,M_{\odot}$ are transfered to the neutron star -- enough to create a mildly
($>10\,{\rm ms}$) spun-up millisecond pulsar, see Section~\ref{sec:spinup}.\\
Binaries with either more massive donor stars ($>5\,M_{\odot}$) or initial periods in excess of 10~days will undergo
dynamically unstable RLO (see Paper~I, and references therein) leading to a common envelope and spiral-in.
The outcome of the in-spiral is probably a merger for Case~B RLO, since for these stars the (absolute) binding energy
of the envelope is too large to allow for its ejection -- except if accretion luminosity of the neutron star
can be efficiently converted into kinetic energy of the envelope. 

\subsection{IMXBs with Case~C RLO and a common envelope}
Donor stars in systems with very wide orbits ($P_{\rm orb}\simeq 10^2-10^3$~days) prior to the mass-transfer phase
develop a deep convective envelope as they become giant stars before filling their Roche-lobe.
The response to mass loss for these stars is therefore expansion 
which causes the stars to overfill their Roche-lobe even more.
To exacerbate this problem, binaries also shrink in size if mass transfer occurs from a
donor star somewhat more massive 
than the accreting neutron star. This causes further
overfilling of the donor star Roche-lobe resulting in enhanced mass loss etc.
This situation is a vicious circle that leads to a dynamically unstable, runaway mass transfer and the formation of a
common envelope (CE) followed by a spiral-in phase, e.g. \citet{pac76}, \citet{il93}, \citet{ijc+12}.
Whereas donor stars which initiate RLO during hydrogen shell burning still have relatively tightly bound envelopes, these stars here which
initiate RLO on the AGB have only weakly bound envelopes \citep{hpe94,dt00}. Hence, these donor stars can survive the spiral-in phase
of the neutron star without merging.
To leave a {CO}~WD remnant from such wide-orbit systems the ZAMS masses of the progenitor stars (the donor stars of the IMXBs) must 
be $2.2-6\,M_{\odot}$ (Paper~I), depending on the assumed amount of convective core-overshooting in the stellar models.

\subsection{BMSPs with a {CO}~WD from LMXBs} \label{subsec:COWD-LMXB}
In addition to the formation channels from IMXBs, pulsars with a {CO}~WD companion
can, in rare cases, be produced from late Case~B RLO in an LMXB system -- e.g. see Table~A1 in \citet{ts99}.
The outcome is an extremely wide-orbit pulsar system ($P_{\rm orb}>1000\,{\rm days}$)
with a $0.47-0.67\,M_{\odot}$ {CO}~WD and a slowly spinning pulsar.
The pulsar B0820+02 ($P=0.86\,{\rm sec}$, $P_{\rm orb}=1232\,{\rm days}$) is an example of a system
which followed this formation channel (cf. Section~\ref{subsec:corbet}).
It is an interesting fact that {CO}~WDs produced from donor stars in IMXBs can be less massive than
{CO}~WDs produced in such LMXBs.
In low-mass stars ($\le 2.2\,M_{\odot}$) helium is ignited in a flash under degenerate conditions and 
{CO}~WDs produced from these stars in LMXBs have minimum masses of $0.47\,M_{\odot}$, whereas in cores of intermediate-mass stars 
helium ignites non-degenerately and {CO}~WDs from IMXBs can be made with masses down to $0.33\,M_{\odot}$ \citep{kw90,tvs00,prp02}.

\subsection{BMSPs with a {CO}~WD from AIC?}
Under certain circumstances is may be possible for an accreting {O}-{Ne}-{Mg}~WD to reach the Chandrasekhar mass limit and 
thereby implode to form a neutron star -- the so-called accretion-induced collapse, AIC \citep{mnys80,nom84,cil90}. A neutron star formed
this way may leave behind a radio pulsar. The necessary conditions for a pulsar formed this way to further accrete from
the companion star and leaving behind, for example a BMSP with a {CO}~WD companion, is investigated 
in a separate paper (Tauris~et~al., in preparation). However, the recycling process,
which is the main topic in this paper, remains the same. 

\begin{figure*}
\begin{center}
  \includegraphics[width=0.60\textwidth, angle=-90]{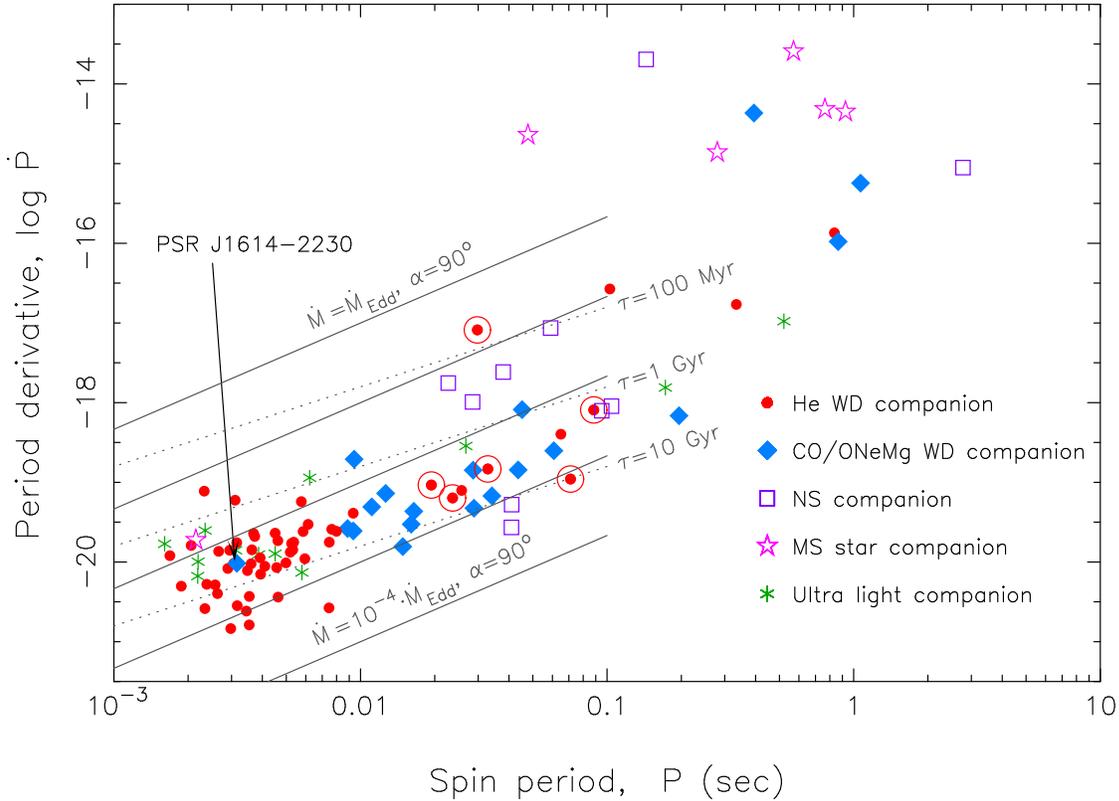}
  \caption[]{
    Distribution in the $P\dot{P}$--diagram of 103 binary radio pulsars in the Galactic disk.
    The five solid lines indicate theoretical spin-up lines using equation~(\ref{eq:spinuplinefit2})
    for a $1.4\,M_{\odot}$ pulsar accreting with a rate 
    of $\dot{M}=\dot{M}_{\rm Edd}$, $0.1\,\dot{M}_{\rm Edd}$, 
    $0.01\,\dot{M}_{\rm Edd}$, $0.001\,\dot{M}$ and $\dot{M}=10^{-4}\,\dot{M}_{\rm Edd}$, respectively (top to bottom).
    Also shown as dotted lines are characteristic ages
    of 100~Myr, 1~Gyr and 10~Gyr, respectively.
    The location of \psr\ is peculiar in the sense that it has a massive {CO}~WD companion (blue diamond) in a region where
    all other pulsars have a low-mass {He}~WD companion (red filled circles).
    The six pulsars marked with circles may have evolved from IMXBs and some of these are potential candidates for having a CO~WD companion
    -- see Section~\ref{subsec:corbet}. Data taken from the {\it ATNF Pulsar Catalogue},
    http://www.atnf.csiro.au/research/pulsar/psrcat/ \citep{mhth05}, in May~2012.
    }
\label{fig:ppdot}
\end{center}
\end{figure*}

%%%%%%%%%%%%%%%%%%%%%%%%%%%%%%%%%%%%%%%%%%%%%%%%%%%%%%%%%%%%%%%%%%%%%%%%%%%%%%%%

\section{Observational characteristics of BMSPs with {CO}~WD companions}\label{sec:observations}
Before looking into the details of recycling a pulsar in an {X}-ray binary, we first
summarize the observational properties of the end products, the BMSPs. This is important for our subsequent discussions
and understanding of the theoretical modelling which constitutes the core of our work. 
Furthermore, we shall investigate if and how the BMSPs with {CO}~WD companions 
can be distinguished from the large population of BMSPs with {He}~WD companions. For this purpose we
make use of the $P\dot{P}$--diagram, the Corbet~diagram and information of orbital eccentricities.
 
\subsection{The $P\dot{P}$--diagram}\label{subsec:ppdot}
In order to understand the formation and evolution of recycled pulsars 
we have plotted in the $P\dot{P}$--diagram in Fig.~\ref{fig:ppdot} the distribution of 103 binary radio pulsars 
located in the Galactic disk.
The various companion types are marked with different symbols (see also Table~\ref{table:companions}).
The majority of the companion stars are helium white dwarfs ({He}~WDs).
However, there is also a significant population of the more massive {CO} and {ONeMg}~WDs,  
as well as pulsars with another neutron star as companion. In some systems the companion is still a main sequence star. In other
(very tight) systems the companion has evolved into a semi-degenerate substellar remnant ($<0.08\,M_{\odot}$) 
-- possibly aided by evaporation caused by the illumination from the pulsar following an 
ultra-compact LMXB phase \citep{pod91,db03,vvp05,bbb+11,vnv+12}.\\ 
Pulsar systems found in globular clusters are not useful as probes of stellar evolution and binary interactions since their
location in a dense environment causes frequent exchange collisions whereby the orbital parameters are perturbed or the
companion star is replaced with another star and thus information
is lost regarding the mass transfer and spin-up of the pulsar. These pulsars could even have undergone accretion
events from two or more different companions.
%-------------------------------------------------------------------------------
\begin{table}
\center
\caption{Binary radio pulsars and their companions.
         Data taken from the {\it ATNF Pulsar Catalogue} in May~2012.}
\begin{tabular}{lr}
\hline {Population} & {Number} \\ 
\hline 
\noalign{\smallskip} 
            All known binary pulsars: & 186 \\
\noalign{\smallskip} 
            $\quad$Galactic disk            & 110 \\
            $\quad$Extra galactic (SMC)     & 1 \\
            $\quad$Globular clusters        & 75 \\
\hline 
\noalign{\smallskip} 
            Galactic disk$^*$ binary pulsars with measured $\dot{P}$:   & 103 \\
\noalign{\smallskip} 
            Companion star: &    \\
\noalign{\smallskip} 
            $\quad${He}~white dwarf                           & 56 \\
            $\quad${CO/ONeMg}~white dwarf                     & 19 \\
            $\quad$Neutron star                               & 10 \\
            $\quad$Main sequence star$^*$                     &  6 \\
            $\quad$Ultra~light (semi)deg. dwarf, or planet(s) & 12 \\
\noalign{\smallskip} 
\hline
\end{tabular}
\begin{flushleft}
  $^*$ Including PSR~J0045$-$7319 which is located in the Small Magellanic Cloud 
       and has a  $B1${\Large \romannumeral 5} companion \citep{kjb+94}.
\end{flushleft}
\label{table:companions}
\end{table}
%-------------------------------------------------------------------------------
%-------------------------------------------------------------------------------
\begin{table}
\center
\caption{Binary radio pulsars which are likely to have evolved from IMXB systems. 
         Data taken from the {\it ATNF Pulsar Catalogue} in May~2012.}
\begin{tabular}{lrrrrl}
\hline {PSR} & {$P_{\rm orb}$} & {$P$} & {$\dot{P}$}  & {ecc}        & {$M_{\rm WD}$} \\ 
        {}   & {days}          & {ms}  & {$10^{-19}$} & {$10^{-4}$}  & {$M_{\odot}$} \\ 
\hline 
\noalign{\smallskip} 
 J1952+2630   &    0.392 & 20.7 & n/a   &  n/a    & 1.13 \\
 J1757$-$5322 &    0.453 & 8.87 & 0.263 &  0.0402 & 0.67 \\
 J1802$-$2124 &    0.699 & 12.6 & 0.726 &  0.0247 & 0.78$^*$ \\
 B0655+64     &    1.03  & 196  & 6.85  &  0.0750 & 0.80 \\
 J1435$-$6100 &    1.35  & 9.35 & 0.245 &  0.105  & 1.08 \\
 J1439$-$5501 &    2.12  & 28.6 & 1.42  &  0.499  & 1.30$^*$ \\
 J1528$-$3146 &    3.18  & 60.8 & 2.49  &  2.13   & 1.15 \\
 J1157$-$5112 &    3.51  & 43.6 & 1.43  &  4.02   & 1.30$^*$ \\
 J1337$-$6423 &    4.79  & 9.42 & 1.95  &  0.197  & 0.95 \\ 
 J1603$-$7202 &    6.31  & 14.8 & 0.156 &  0.0928 & 0.34 \\
 J2145$-$0750 &    6.84  & 16.1 & 0.298 &  0.193  & 0.50 \\
 J1022+1001   &    7.81  & 16.5 & 0.433 &  0.970  & 0.85 \\
 J0621+1002   &    8.32  & 28.9 & 0.473 &  24.6   & 0.67$^*$ \\
 J1614$-$2230 &    8.69  & 3.15 & 0.0962&  0.0130 & 0.50$^*$ \\
 J1454$-$5846 &   12.4   & 45.2 & 8.17  &  19.0   & 1.05 \\
 J0900$-$3144 &   18.7   & 11.1 & 0.491 &  0.0103 & 0.42 \\
 J1420$-$5625 &   40.3   & 34.1 & 0.675 &  35.0   & 0.44 \\
\hline 
 J1141$-$6545$^a$ &    0.198 & 394  & 4307  &  1719   & 1.02$^*$ \\
 B2303+46$^a$     &   12.3   & 1066 & 569   &  6584   & 1.30$^*$ \\
 B0820+02$^b$     & 1232     & 865  & 105   &  118.7  & $>0.52$ \\
\hline 
 J1622$-$6617$^c$ &  1.64 & 23.6 &  0.636 & n/a   & 0.11\\
 J1232$-$6501$^c$ &  1.86 & 88.3 &  8.11  & 1.09  & 0.17\\
 J1745$-$0952$^c$ &  4.94 & 19.4 &  0.925 & 0.0985& 0.13\\
 J1841+0130$^c$   & 10.5  & 29.8 &  81.7  & 0.819 & 0.11\\
 J1904+0412$^c$   & 14.9  & 71.1 &  1.10  & 2.20  & 0.26\\
 J1810$-$2005$^c$ & 15.0  & 32.8 &  1.47  & 0.192 & 0.33\\
%J1822$-$0848$^d$ & 287   & 835  &  1350  & 590   & 0.38\\
\noalign{\smallskip} 
\hline
\end{tabular}
\begin{flushleft}
  $^*$ Mass obtained from timing. Other masses are median masses
       (i.e. calculated for $i=60^{\circ}$ and an assumed $M_{\rm NS}=1.35\,M_{\odot}$).\\ 
  $^a$ A non-recycled pulsar formed {\it after} its WD companion \citep{ts00}.\\ 
  $^b$ This system descends from a very wide-orbit LMXB (see text).\\ 
  $^c$ CO~WD candidate or {He}~WD evolving from an IMXB.\\ 
\end{flushleft}
\label{table:COcompanions}
\end{table}
%-------------------------------------------------------------------------------

The location of \psr\ in the $P\dot{P}$--diagram is quite unusual for a BMSP with a {CO}~WD companion.
It is located in an area which is otherwise only populated by BMSPs with {He}~WD companions. This 
is important as it proves for the first time
that efficient, full recycling (i.e. $P\le 8\,{\rm ms}$) 
of a pulsar is also possible in {X}-ray systems leading to BMSPs with {CO}~WD companions.
In general, the BMSPs with CO~WD companions are seen to have large values of both $P$ and $\dot{P}$ compared to
BMSPs with {He}~WD companions. This trend can be understood since the former systems evolve from IMXBs which often
have short lasting RLO and thus inefficient spin-up -- \psr\ being the only known exception (this system evolved
from an IMXB via Case~A RLO, see Paper~I).
\begin{figure}
\begin{center}
  \includegraphics[width=0.35\textwidth, angle=-90]{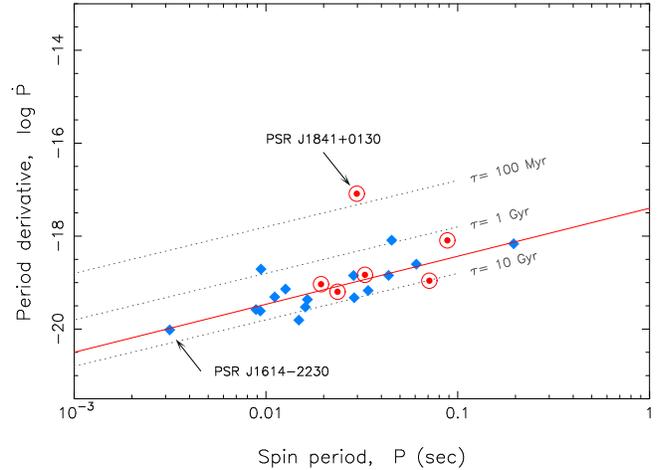}
  \caption[]{The observed recycled pulsars with {CO} WD companions (diamonds) seem to cluster somewhat along a straight line in the
             $P\dot{P}$--diagram. The full line is a linear regression fit to all these pulsars and the dotted
             lines represent constant characteristic ages. Other pulsars which are also candidates for having
             an IMXB origin are shown as dots in a circle (cf. Section~\ref{subsec:corbet}).
             PSR~J1841+0130 is a young example of such a candidate -- see text. 
    }
\label{fig:line}
\end{center}
\end{figure}

In Table~\ref{table:COcompanions} we list all binary radio pulsars with {CO}~WD 
(or {ONeMg}~WD) companions  -- see also \citet{hfs+04}. At the 
bottom of the Table we list an additional number of sources which we identify as {CO}~WD candidate systems 
(alternatively, these candidates may be {He}~WDs evolving from IMXBs).
The three slowest rotating of the 20 pulsars with a {CO}~WD companion are not recycled -- either because they were formed {\it after}
the WD \citep[][]{ts00}, or because they formed in a very wide orbit (cf. Section~\ref{subsec:COWD-LMXB}).
Of the remaining 17 recycled pulsars with a {CO}~WD companion, 16 have a measured value of $\dot{P}$. 
These are plotted in Fig.~\ref{fig:line} together with those other systems which may also have evolved from IMXB systems
and thus possibly host a {CO}~WD companion too (cf. Section~\ref{subsec:corbet}).
With the exception of PSR~J1841+0130 (which is young, see Section~\ref{subsubsec:1841}), these pulsars are 
distributed in a somewhat more linear manner in the $P\dot{P}$--diagram compared to
BMSPs with a {He}~WD companion. 

\subsection{The Corbet diagram - six unusual pulsars}\label{subsec:corbet}
\begin{figure}
\begin{center}
  \includegraphics[width=0.40\textwidth, angle=0]{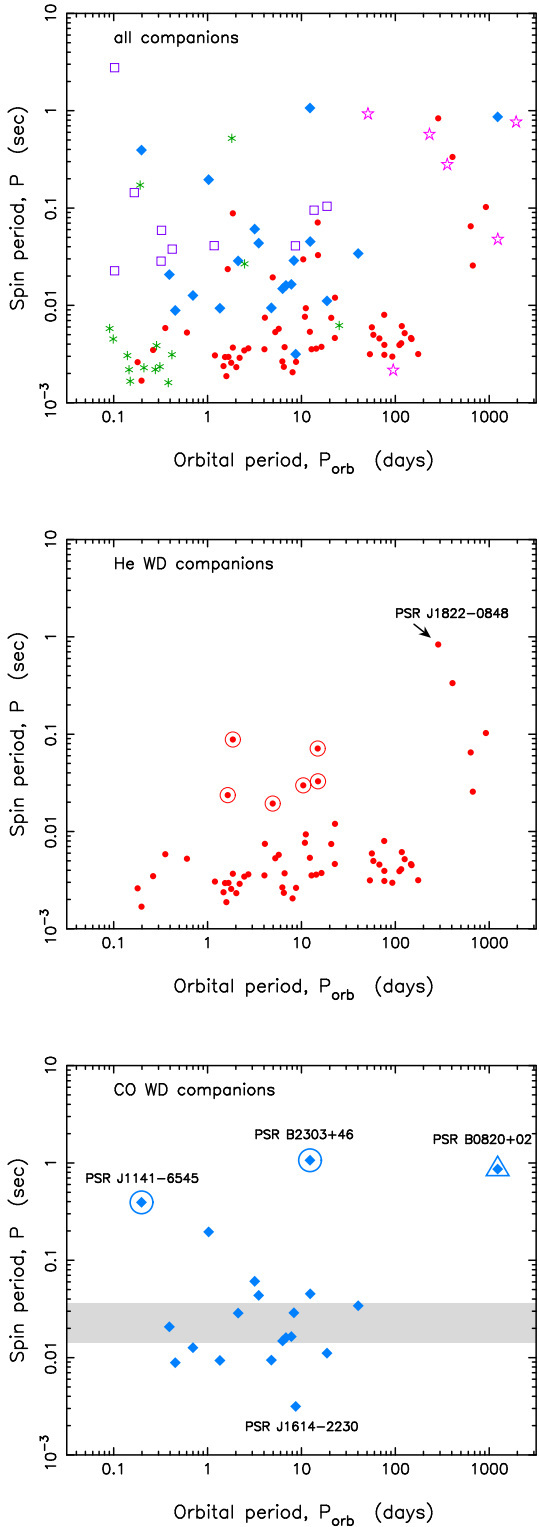}
  \caption[]{
    All binary radio pulsars in the Galactic disk plotted in a Corbet~diagram (top), see Fig.~\ref{fig:ppdot} for an explanation of the symbols plotted.
    Due to their missing $\dot{P}$ measurements 8 of these pulsars were not shown in Fig.~\ref{fig:ppdot}.  
    The central and bottom panels show the distribution of pulsars with {He}~WD (60 systems) 
    and {CO}~WD companions (20 systems), respectively. In the bottom panel the grey shaded region marks the
    range of spin periods obtained from one of our model calculations of Case~BB RLO in a post-CE binary (Fig.~\ref{fig:HeNS}). 
    See text for further discussion.
    }
\label{fig:corbet3}
\end{center}
\end{figure}
\begin{figure}
\begin{center}
  \includegraphics[width=0.35\textwidth, angle=-90]{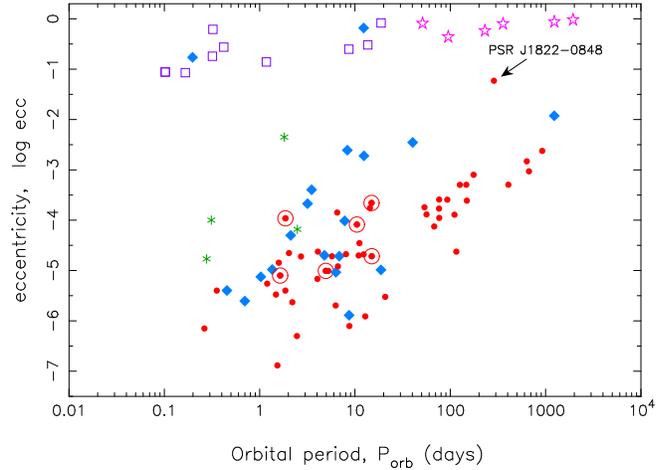}
  \caption[]{
    Eccentricity as a function of orbital period for binary radio pulsars in the Galactic disk.
    The various symbols are defined in Fig.~\ref{fig:ppdot}.
    We notice PSR~J1822$-$0848 which has an eccentricity two orders of magnitude larger than other wide-orbit
    binary pulsars with a {He}~WD companion.
    }
\label{fig:ecc}
\end{center}
\end{figure}
 
Plotting BMSPs in the Corbet~diagram \citep{cor84} could lead to potential clues about their formation history.
In the top panel of Fig.~\ref{fig:corbet3} we have shown the distribution of all binary pulsars.
Little can be learnt from this plot when considering the entire ensemble of binary pulsars
as one population. In the central plot, however, we display only those binary pulsars which have 
a {He}~WD companion. The distribution of pulsars in this diagram shows some interesting features. 
The six pulsars
marked with a circle are noticeable for having unusual slow spin periods compared to other pulsars with similar orbital periods,
see Table~\ref{table:COcompanions} for further details.
This fact hints an inefficient and shorter spin-up phase compared to the BMSPS with fast spin.
\citet{tau11} suggested that these pulsars may perhaps have formed via the accretion induced collapse (AIC)
of a WD. Here, we suggest instead that these pulsars (which all have $P_{\rm orb}\ge 1\;{\rm day}$) may have formed in IMXBs. In that case the companions
are likely to be {CO}~WDs, although {He}~WD companions may also form from IMXBs \citep{tvs00,prp02}.
Similar ideas of an IMXB origin have been proposed by \citet{clm+01} and \citet{li02}. The latter author argued that accretion disk
instabilities in IMXBs may explain the slow spin periods and the high $\dot{P}$ values of these pulsars.    
The pulsars with $P_{\rm orb}>200\;{\rm days}$ all have slow spin periods -- see Section~\ref{subsec:wLMXBs} for
a discussion on these systems. 
The bottom panel shows the binary pulsars with {CO}~WD companions.
The main thing to notice is that these systems have $P_{\rm orb} \le 40^{\rm d}$ and slower
pulsar spin periods compared to BMSPs with {He}~WD companions, as first pointed out by \citet{cam96}.
The slower spin of pulsars in this population can be understood well which is discussed further
in Section~\ref{sec:obs-spin}.
A few pulsars marked with a circle or a triangle stand out from the rest of the population. These are
PSR~J1141$-$6545 and PSR~B2303+46 where the neutron star formed after the white dwarf companion \citep{ts00} and
therefore these pulsars remain non-recycled with slow spin periods, and PSR~B0820+02 which formed from
an extremely wide-orbit LMXB with short lasting and possibly inefficient RLO 
(see Sections~\ref{subsec:COWD-LMXB} and \ref{subsec:wLMXBs}).

\subsubsection{PSR~J1841+0130}\label{subsubsec:1841wd}
PSR~J1841+0130 \citep[discovered by ][]{lfl+06} is one of the six pulsars which we note to have an offset location in the Corbet diagram
when comparing to BMSPs with {He}~WD companions (Fig.~\ref{fig:corbet3}, central panel). 
It has a slow spin period of 29~ms and an orbital period, $P_{\rm orb}=10.5^{\rm d}$. 
The majority of other BMSPs with similar $P_{\rm orb}$ near 10~days have much faster spin periods 
between $2-10\;{\rm ms}$.
The slow spin period of PSR~J1841+0130 can be explained if it originated from an IMXB system rather than an LMXB system. 
The mass function of this pulsar, $f=4.22\times 10^{-4}\,M_{\odot}$ suggests
that its companion star has a mass of $0.11\,M_{\odot}$ for an orbital inclination angle of $60^\circ$. 
If the companion star is a {CO}~WD formed in an IMXB it must have a mass $\ge 0.33\,M_{\odot}$
(cf.~Section~\ref{subsec:COWD-LMXB}). 
This would require a small inclination angle, $i\le 20^\circ$ (depending on the neutron star mass). The probability 
for this to happen from a distribution of random orbital orientations
is about 6~per~cent. 
The situation is similar in PSR~J1622$-$6617 which is another candidate systems to have evolved from an IMXB.
However, IMXBs can also leave behind {He}~WDs which have somewhat smaller masses.
Alternatively, if PSR~J1841+0130 originated from an LMXB system then according to the $(M_{\rm WD},\,P_{\rm orb}$)-relation,
e.g. \citet{ts99}, we would expect $i\approx 25^\circ$ to obtain the predicted {He}~WD mass of $0.26\pm 0.01\,M_{\odot}$.
(The probability for this is about 9~per~cent.)
Hence, even in this case the inclination angle would be rather low.
As discussed in Sections~\ref{subsec:spinupline} and \ref{subsubsec:1841}, PSR~J1841+0130 is also interesting for its ability to 
constrain spin-up physics and for its young age.

\subsection{The orbital eccentricity}\label{subsec:ecc}
Another fossil record of binary evolution is the orbital eccentricity \citep{phi92,pk94}. In Fig.~\ref{fig:ecc}
we have plotted the eccentricity as a function of orbital period. 
The BMSPs with {He}~WDs have in general somewhat lower eccentricities than systems with {CO}~WDs. 
The spread is large for both 
of the two populations which also overlap. Among the systems with {CO}~WD companions it is not
surprising that \psr\ has the lowest eccentricity since we have demonstrated (Paper~I) that this system formed
via a relatively long phase of stable RLO. It is interesting to notice that the four BMSPs with {CO}~WDs and 
$P_{\rm orb}<2\,{\rm days}$ all have quite small eccentricities $\le 10^{-5}$ even though they are believed to have formed via a CE.

\subsubsection{PSR~J1822$-$0848}\label{subsubsec:1822}
PSR~J1822$-$0848 \citep{lfl+06} has an orbital period of 287~days and an eccentricity, $ecc=0.059$. 
From Fig.~\ref{fig:ecc} it is seen that this eccentricity is two orders of magnitude larger than
that of other pulsars in similar wide-orbit systems. This fact, combined with its very slow spin period of 835~ms 
(see the central panel in Fig.~\ref{fig:corbet3}) and its high value of $\dot{P}=1.35\times 10^{-16}$, hints
that this pulsar may not have a {He}~WD companion star. It is possible it belongs to the same class as
PSR~B0820+02 which is in a wide~orbit with a {CO}~WD (see Section~\ref{subsec:COWD-LMXB}). However, in that case
it needs to be explained why some pulsars have {He}~WD companions in even larger orbits compared to PSR~J1822$-$0848.
Alternatively, one could speculate that PSR~J1822$-$0848 experienced weak spiral-in from an almost unbound common envelope 
of an AGB star. This could explain its non-recycled characteristics and why it has $P_{\rm orb}<1000\;{\rm days}$.\\
Although, it has a relatively large eccentricity of 0.059 this value is still far too small to suggest that PSR~J1822$-$0848 
was born after the WD, like in the case of PSR~1141$-$6545 and PSR~B2303+46 \citep{ts00}.
Even in the unrealistic case of a collapse of a naked core with no envelope, and 
without a kick, the present eccentricity of 0.059 for such a
wide orbit would require a maximum instant mass loss of only $\sim \! 0.10\,M_{\odot}$ -- a value 
which is even less than the release of the gravitational binding energy during the core collapse. Thus this scenario is not possible and
we therefore conclude that PSR~J1822$-$0848 was formed before its WD companion, like 99~per~cent of all detected binary pulsars
with WD companions.

%%%%%%%%%%%%%%%%%%%%%%%%%%%%%%%%%%%%%%%%%%%%%%%%%%%%%%%%%%%%%%%%%%%%%%%%%%%%%%%%

\section{Recycling the pulsar}\label{sec:spinup}
We now proceed with a general examination 
of the recycling process of pulsars. 
We begin by analysing the concept of spin-up lines in the $P\dot{P}$--diagram 
and derive an analytic expression for the amount of accreted mass needed to spin~up a pulsar
to a given equilibrium spin period.
We use standard theory for deriving these expressions,
but include it here to present a coherent and detailed derivation needed for our
purposes. Furthermore, we apply the \citet{spi06} solution to the pulsar spin-down torque which combines
the effect of a plasma current in the magnetosphere with the magnetic dipole model. 
The physics of the disk--magnetosphere interaction is still not known in detail.
The interplay between the neutron star magnetic field and the conducting plasma in the accretion disk
is a rather complex process.
For further details of the accretion physics we 
refer to \citet{pr72,lpp73,do73,gl79a,gl79b,st83,gl92,st93,ccm+98,fkr02,rfs04,bsvg09,ds10,ib12} and references therein.

\subsection{The accretion disk}\label{subsec:accdisk}
The accreted gas from a binary companion possess large specific angular momentum. For this reason the flow of gas onto 
the neutron star is not spherical, but leads to the formation of an accretion disk where excess
angular momentum is transported outwards by (turbulent-enhanced) viscous stresses, e.g. \citet{st83,fkr02}.
Depending on the mass-transfer rate, the opacity of the accreted material and the temperature of the disk, 
the geometric shape and flow of the material may take a variety of
forms (thick disk, thin thick, slim disk, torus-like, ADAF). Popular models of the inner disk \citep{gl92}
include optically thin/thick disks which can be either gas (GPD) or radiation pressure dominated (RPD).
The exact expression for the spin-up line in the $P\dot{P}$--diagram also depends on the assumed model for the 
inner disk -- mainly as a result of 
the magnetosphere boundary which depends on the characteristics of the inner disk.
Close to the neutron star surface the magnetic field is strong enough that
the magnetic stresses truncate the Keplerian disk, and the plasma is channeled along field lines to accrete 
on to the surface of the neutron star. At the inner edge of the disk the magnetic field
interacts directly with the disk material over some finite region.
The physics of this transition zone from Keplerian disk to magnetospheric flow is important and
determines the angular momentum exchange from the differential rotation between the disk
and the neutron star.\\
Interestingly enough, it seems to be the case that the resultant accretion torque, acting on the neutron star, calculated using 
detailed models of the disk--magnetosphere interaction does not deviate much from simple
expressions assuming idealized, spherical accretion and newtonian dynamics. 
For example, it has been pointed out by \citet{gl92} as a fortuitous coincidence that the
equilibrium spin period calculated under simple assumptions of spherical flow resembles the more detailed
models of an optically thick, gas pressure dominated inner accretion disk.
Hence, as a starting point this allows us to expand on standard prescriptions in the literature with the aims of: 
1) performing a more careful analysis of the concept of a spin-up line, 2) deriving
an analytic expression for the mass needed to spin~up a given observed millisecond pulsar, and 3)
understanding the effects on the spinning neutron star when the donor star decouples from its Roche-lobe. 
As we shall discuss further in Section~\ref{sec:RLDP}, the latter issue depends on the location of the magnetospheric boundary 
(or the inner edge of the accretion disk) relative to the 
corotation radius and the light-cylinder radius of the pulsar. 
All stellar parameters listed below will refer
to the neutron star unless explicitly stated otherwise.

\subsection{The accretion torque -- the basics}\label{subsec:torque}
The mass transfered from the donor star carries with it angular momentum which eventually
spins up the rotating neutron star once its surface magnetic flux density, $B$
is low enough to allow for efficient accretion, i.e. following initial phases 
where either the magneto-dipole radiation pressure dominates or  
propeller effects are at work.\\
The accretion torque acting on the spinning neutron star has a contribution from both material stress (dominant term),
magnetic stress and viscous stress, and is given by:
$N = \dot{J}_\star \equiv (d/dt) (I\Omega _\star)$, 
where $J_\star$ is the pulsar spin angular momentum, $\Omega _\star$ is its angular velocity and
 $I\approx 1$--$2\times 10^{45}\;{\rm g~cm}^2$ is its 
moment of inertia. 
The exchange of angular momentum ($\vec{J}=\vec{r}\times\vec{p}$) at the magnetospheric boundary eventually leads to a
gain of neutron star spin angular momentum which can approximately be expressed as:   
\begin{equation}
  \Delta J_\star = \sqrt{GMr_{\rm A}}\,\Delta M \,\xi
  \label{eq:deltaJ}
\end{equation}
where $\xi \simeq 1$ is a numerical factor which depends on the flow pattern \citep{gl79b,gl92},
$\Delta M = \dot{M}\cdot\Delta t$ is the amount of mass accreted in a time interval $\Delta t$
with average mass accretion rate $\dot{M}$ and 
\begin{eqnarray}\label{eq:alfven}
   r_{\rm A} & \simeq & \left( \frac{B^2\,R^6}{\dot{M}\sqrt{2GM}} \right) ^{2/7}  \\ \nonumber
             & \simeq & 22\,{\rm km}\quad B_8^{4/7}\left(\frac{\dot{M}}{0.1\,\dot{M}_{\rm Edd}}\right) ^{-2/7} \left(\frac{M}{1.4\,M_{\odot}}\right) ^{-5/7}
\end{eqnarray}
is the Alfv\'en radius defined as the location where the magnetic energy density will
begin to control the flow of matter (i.e. where the incoming material couples to the magnetic field lines
and co-rotate with the neutron star magnetosphere). Here $B$ is the surface magnetic flux density, $R$ is the neutron star radius,
$M$ is the neutron star mass (see relation between $R$ and $M$ further below) and $B_8$ is $B$ in units of $10^8$~Gauss.
A typical value for the Alfv\'en radius in accreting {X}-ray millisecond pulsars (AXMSPs), 
obtained from $B\sim\!10^8\,{\rm G}$ and $\dot{M}\sim\!0.01\,\dot{M}_{\rm Edd}$, is $\sim\!40\,{\rm km}$
corresponding to about $3\,R$.
The expression above is found by equating the magnetic energy density ($B^2/8\pi$)
to the ram pressure of the incoming matter and using the continuity equation \citep[e.g.][]{pr72}. Furthermore, it assumes a scaling
with distance, $r$ of the far-field strength of the dipole magnetic moment, ${\mu}$ as: $B(r)\propto {\mu}/r^3$ (i.e. disregarding
poorly known effects such as magnetic screening \citep{vas79}).
A more detailed estimation of the location of the inner edge of the disk, i.e. the coupling radius or magnetospheric
boundary, is given by: $r_{\rm mag}=\phi\cdot r_A$, where $\phi$ is $0.5-1.4$ \citep{gl92,wan97,ds10}.

\subsubsection{The surface B-field strength of recycled radio pulsars}\label{subsubsec:B}
Before we proceed we need an expression for $B$. 
One can estimate the B-field of recycled MSPs based on their observed spin period, $P$ and its time derivative $\dot{P}$.
The usual assumption is to apply the vacuum magnetic dipole model in which
the rate of rotational energy loss ($\dot{E}_{\rm rot}=I\Omega\dot{\Omega})$ is equated to the energy-loss rate
caused by emission of dipole waves (with a frequency equal to the spin frequency of the pulsar) due to an inclined axis of 
the magnetic dipole field with respect to
the rotation axis of the pulsar:
\begin{equation} 
  \dot{E}_{\rm dipole}=(-2/3c^3)|\ddot{\mu}|^2
  \label{eq:dipole}
\end{equation} 
The result is:
\begin{eqnarray}\label{eq:B} 
     B_{\rm dipole}   & = & \sqrt{\frac{3c^3IP\dot{P}}{8\pi ^2 R^6}}\;\frac{1}{\sin\alpha}\\ \nonumber 
         & \simeq & 1.6\times 10^{19}\,G\quad\sqrt{P\dot{P}}\;\left( \frac{M}{1.4\,M_{\odot}}\right) ^{3/2}\frac{1}{\sin\alpha}
\end{eqnarray}
where the magnetic inclination angle is $0<\alpha\le 90^{\circ}$.
This is the standard equation for evaluating the B-field of a radio pulsar. Our numerical scaling constant
differs by a factor of a few from the conventional one:
$B=3.2\times 10^{19}\,G\;\sqrt{P\dot{P}}$, which assumes $R=10\;{\rm km}$ and $I=10^{45}\;{\rm g\,cm^2}$, 
both of which we believe are slightly underestimated values (our assumed relations
between $I$, $M$ and $R$ are discussed in Section~\ref{subsubsec:eddlimit}).
Note also that some descriptions in the literature apply the polar B-field strength ($B_p=2\,B$) rather than
the equatorial B-field strength used here.\\ 
It is important to realize that the above expression does not include the rotational energy loss obtained when considering the spin-down torque
caused by the $\vec{j}\times\vec{B}$ force
exerted by the plasma current in the magnetosphere \citep[e.g.][]{gj69,spi08}.  
For this reason the vacuum magnetic dipole model does not predict any spin-down torque for an aligned rotator ($\alpha = 0^{\circ}$),
which is not correct. The incompleteness of the vacuum magnetic dipole model was in particular evident after the discovery
of intermittent pulsars by \citet{klo+06} and demonstrated the need for including the plasma term in the spin-down torque.
A combined model was derived by \citet{spi06} and applying his 
relation between $B$ and $\alpha$ we can rewrite the above expression slightly:
\begin{eqnarray}\label{eq:Bspitkovsky} 
     B   & = & \sqrt{\frac{c^3IP\dot{P}}{4\pi ^2 R^6}\;\frac{1}{1+\sin^2\alpha}}\\ \nonumber 
         & \simeq & 1.3\times 10^{19}\,G\quad\sqrt{P\dot{P}}\;\left( \frac{M}{1.4\,M_{\odot}}\right) ^{3/2} \sqrt{\frac{1}{1+\sin^2\alpha}}
\end{eqnarray}  
This new expression leads to smaller estimated values of $B$ by a factor of at least $\sqrt{3}$, or more precisely
$\sqrt{\frac{3}{2}(2+\cot ^2 \alpha)}$, compared to the vacuum dipole model.
As we shall shortly demonstrate, this difference in dependence on $\alpha$ is quite important for the location of
the spin-up line in the $P\dot{P}$--diagram.

\subsubsection{The Eddington accretion limit}\label{subsubsec:eddlimit}
The Eddington accretion limit is given by:
\begin{eqnarray} 
  \label{eq:Medd}
  \dot{M}_{\rm Edd} & = & \frac{4\pi c\, m_p}{\sigma _{\rm T}}\,R\,\mu _{\rm e} \\ \nonumber
                    & \simeq & 3.0\times 10^{-8}\,M_{\odot}\,{\rm yr}^{-1}\quad\!\! R_{13}\left(\frac{1.3}{1+X}\right)
\end{eqnarray}
where $c$ is the speed of light in vacuum, $m_p$ is the proton mass, $\sigma _{\rm T}$ is the Thomson scattering
cross section, $\mu _{\rm e}=2/(1+{X})$ is the mean molecular weight per electron which depends on the 
hydrogen mass fraction, {X} of the accreted material, and $R_{13}$ is the neutron star radius in units of 13~km. 
The expression is found by equating the outward radiation pressure
to the gravitational force per unit area acting on the nucleons of the accreted plasma.
%The radiation pressure originates from photons, released via heated matter falling down 
%the deep gravitational potential well of the neutron star, which
%scatter on the electrons in the incoming ionised plasma. The released luminosity is given by: $L=G\dot{M}M/R$. 
In general, luminosity is generated from both nuclear burning at the neutron star surface as well as 
from the release of gravitational binding energy of the accreted material, 
i.e. $L=(\epsilon_{\rm nuc}+\epsilon_{\rm acc})\,\dot{M}$. However, for accreting neutron stars
$\epsilon_{\rm nuc} \ll \epsilon_{\rm acc}$
and thus we have neglected the contribution from nuclear processing.
%The efficiency
%%of radiative emission in units of rest mass energy is 10--20\% depending on neutron star mass and its equation-of-state.
%The canonical value used in the literature for the Eddington accretion rate is $1.5\times 10^{-8}\,M_{\odot} {\rm yr}^{-1}$ and
%is found for a neutron star radius of 10~km and assuming accretion of pure hydrogen ({X}=1). Since the equation-of-state for
%neutron stars is found to be rather stiff (especially known now as a result of observations of \psr) neutron star radii are probably 
%a bit larger, $R=12$--$15\,{\rm km}$ \citep{dpr+10}. Furthermore, actual values for the hydrogen abundance of the accreted
%matter range between ${X}=0.7$--$0.0$ (from solar abundance material to pure helium, respectively) 
%which also increases the value of $\dot{M}_{\rm Edd}$. Especially towards the end of the RLO the value of
%{X} is quite low (in our Case~A scenario ${X}=0.08$ near the end of phase~AB). A better average value to
%use for $\dot{M}_{\rm Edd}$ would therefore be 
%$3.0\times 10^{-8}\,M_{\odot} {\rm yr}^{-1}$. In our code the value of $\dot{M}_{\rm Edd}$ is calculated according
To estimate the neutron star radius
we used a mass-radius exponent following a simple non-relativistic degenerate Fermi-gas polytrope
($R\propto M^{-1/3}$) with a scaling factor such that: $R=15\,(M/M_{\odot})^{-1/3}\,{\rm km}$, calibrated from 
\psr, cf. fig.~3 in \citet{dpr+10}.
In our code the value of $\dot{M}_{\rm Edd}$ is calculated according
to the chemical composition of the transfered material at any given time. 
Note, the value of $\dot{M}_{\rm Edd}$ is only a rough measure since the derivation assumes spherical symmetry,
steady state accretion, Thompson scattering opacity and Newtonian gravity.

\subsection{The spin-up line}\label{subsec:spinupline}
The observed spin evolution of accreting neutron stars often shows rather stochastic variations on a short timescale \citep{bcc+97}.
The reason for the involved dramatic torque reversals is not well known -- see hypotheses listed at the beginning of Section~\ref{sec:RLDP}.
However, the long-term spin rate will eventually tend towards the {\em equilibrium} spin period, $P_{\rm eq}$
-- meaning that the pulsar spins at the same rate as the particles forced to corotate with the $B$-field
at the magnetospheric boundary.
The location of the associated so-called spin-up line for the rejuvenated pulsar in the $P\dot{P}$--diagram can be found by 
considering the equilibrium configuration when the angular velocity of the neutron star is equal to the
Keplerian angular velocity of matter at the magnetospheric boundary where the accreted matter enters the magnetosphere,
i.e. $\Omega _\star = \Omega _{\rm eq} = \omega _c\,\Omega _{\rm K}(r_{\rm mag})$ or:
\begin{eqnarray} 
\label{eq:Peq0}
     P_{\rm eq} & = & 2\pi \sqrt{\frac{r_{\rm mag}^3}{GM}}\,\frac{1}{\omega _c} \\ \nonumber
         & \simeq & 1.40\,{\rm ms}\quad B_8^{6/7}\left(\frac{\dot{M}}{0.1\,\dot{M}_{\rm Edd}}\right) ^{-3/7} \left(\frac{M}{1.4\,M_{\odot}}\right) ^{-5/7} R_{13}^{18/7} 
\end{eqnarray}
where $0.25 < \omega _c \le 1$ is the so-called critical fastness parameter which 
is a measure of when the accretion torque vanishes \citep[depending on the dynamical importance
of the pulsar spin rate and the magnetic pitch angle, ][]{gl79b}.
One must bear in mind that factors which may differ from unity were omitted in the numerical expression above.
In all numerical expressions in this paper we assumed $\sin \alpha = \phi = \xi = \omega _c = 1$.
Actually, the dependence on the neutron star radius, $R$ disappears in
the full analytic formula obtained by
inserting equations~(\ref{eq:alfven}) and (\ref{eq:Bspitkovsky}) into the top expression in
equation~(\ref{eq:Peq0}), and using $r_{\rm mag}=\phi\cdot r_{\rm A}$, which yields:
\begin{equation} 
     P_{\rm eq} = \left( \frac{\pi c^9}{\sqrt{2}\,G^5}\frac{I^3 \dot{P}^3}{M^5\dot{M}^3} \right) ^{1/4} 
                  \,(1+\sin^2\alpha)^{-3/4}\;\,\phi^{21/8}\,\omega _c ^{-7/4}
  \label{eq:Peq}
\end{equation}
Notice, in the above step we needed to
link the B-fields of accreting neutron stars to the B-fields (expressed by $P$ and $\dot{P}$) 
estimated for observed recycled radio pulsars.
This connection can be approximated in the following manner: 
If the radio pulsar after the recycling phase is ``born'' with a spin period, $P_0$ which is somewhat close to 
$P_{\rm eq}$ (i.e. {\it if} the Roche-lobe decoupling phase (RLDP) did not significantly affect the spin period of the pulsar, cf. discussion in Section~\ref{sec:RLDP}),
then we can estimate the location of its magnetosphere
when the source was an AXMSP just prior to the accretion turn-off during the RLDP.
(Further details of our assumptions of the B-fields of accreting neutron stars are given in Section~\ref{subsubsec:Bdecay},
and in Section~\ref{sec:trueages} we discuss the subsequent spin evolution of recycled radio pulsars towards larger periods, $P>P_0$.)

It is often useful to express the time derivative of the spin period as a function the equilibrium spin period, for example
for the purpose of drawing the spin-up line in the $P\dot{P}$--diagram:
\begin{equation}
     \dot{P} = \frac{2^{1/6}G^{5/3}}{\pi ^{1/3} c^3}\frac{\dot{M}M^{5/3}P_{\rm eq}^{4/3}}{I}
                \;\,(1+\sin^{2}\alpha)\;\,\phi^{-7/2}\,\omega_c^{7/3} 
\label{eq:spinupline} 
\end{equation}
Given that $\dot{M}_{\rm Edd}$ is a function of the neutron star radius and using the relation between $M$ and $R$ stated 
in Section~\ref{subsubsec:eddlimit} we need a relation between the
moment of inertia and the mass of the neutron star. According to the equations-of-state studied by
\citet{wkl08} these quantities scale very close to linearly as: $I_{45}\simeq M/M_{\odot}$ (see their fig.~4) where $I_{45}$ is the moment of
inertia in units of $10^{45}\,{\rm g\,cm}^2$. 
Towards the end of the mass-transfer phase the amount of hydrogen in the transfered matter is usually quite small (${X}\le 0.20$). 
The donor star left behind is basically a naked helium core (the proto~WD).
Hence, we can use equation~(\ref{eq:Medd}) and rewrite equation~(\ref{eq:spinupline}) to estimate the
location of the spin-up line for a recycled pulsar in the $P\dot{P}$--diagram only as a function of its mass
and its mass-accretion rate:
\begin{equation} 
     \dot{P} = 3.7\times 10^{-19}\,\;(M/M_{\odot})^{2/3}\,P_{\rm ms}^{4/3} \left(\frac{\dot{M}}{\dot{M}_{\rm Edd}}\right) 
  \label{eq:spinuplinefit2}
\end{equation}
assuming again $\sin \alpha = \phi = \omega _c =1$, and
where $P_{\rm ms}$ is the equilibrium spin period in units of milliseconds.

\begin{figure}
\begin{center}
  \includegraphics[width=0.35\textwidth, angle=-90]{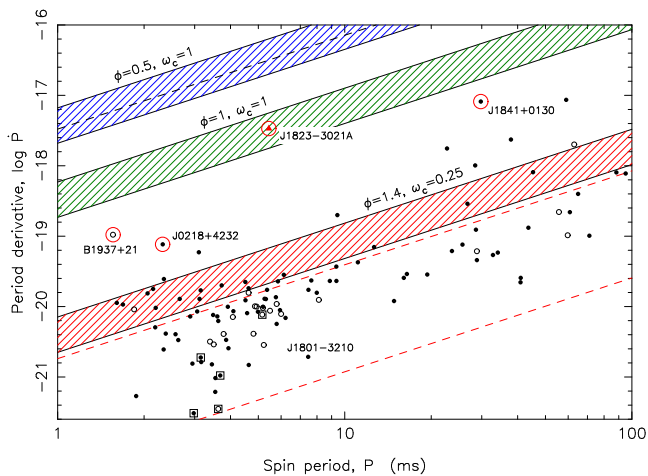}
  \caption[]{
    The spin-up line is shown as three coloured bands depending on the parameters ($\phi$,~$\omega_{\rm c}$).
    In all three cases the spin-up line is calculated assuming accretion at the Eddington limit, $\dot{M}=\dot{M}_{\rm Edd}$
    and applying the Spitkovsky torque formalism. Thus these
    ``lines'' represent upper limits for the given set of parameters. 
    The width of each line (band) results from using a spread in neutron star mass and magnetic inclination angle
    from ($2.0\,M_{\odot}$,~$\alpha=90^{\circ}$) to ($1.0\,M_{\odot}$,~$\alpha=0^{\circ}$),
    upper and lower boundary, respectively. 
    The dashed line within the upper blue band is calculated from ($2.0\,M_{\odot}$,~$\alpha=0^{\circ}$).
    Hence, the band width above this line reflects the dependence on $\alpha$, whereas the band width below
    this line shows the dependence on $M$.
    The two red dashed spin-up lines below the red band are calculated for a $1.4\,M_{\odot}$ neutron star using the 
    vacuum magnetic dipole model for the radio pulsar torque 
    and assuming $\phi =1.4$, $\omega_{\rm c}=0.25$ and $\dot{M}=\dot{M}_{\rm Edd}$ for $\alpha=90^{\circ}$ (upper) and
    $\alpha=10^{\circ}$ (lower).
    The observed distribution of binary and isolated radio pulsars in the Galactic disk 
    (i.e. outside globular clusters) are plotted as filled and open circles, respectively --
    see Fig.~\ref{fig:true_iso_all} for further information. 
    Also plotted is the pulsar J1823$-$3021A which is located in the globular cluster NGC~6624. This pulsar and the
    three other pulsars marked by a red circle are discussed in the text.
    }
\label{fig:spinupline}
\end{center}
\end{figure}
In the literature the spin-up line is almost always plotted without uncertainties.
Furthermore, one should keep in mind the possible effects of the applied accretion disk model
on the location of the spin-up line, cf. Section~\ref{subsec:accdisk}.
In Fig.~\ref{fig:spinupline} we have plotted equation~(\ref{eq:spinupline}) for different values of $\alpha$, $\phi$ and $\omega_{\rm c}$
to illustrate the uncertainties in the applied accretion physics to locate the spin-up line. 
The upper boundary of each band (or ``line'')  
is calculated for a neutron star mass $M=2.0\,M_{\odot}$ and a magnetic inclination angle, $\alpha = 90^{\circ}$.
The lower boundary is calculated for 
$M=1.0\,M_{\odot}$ and $\alpha = 0^{\circ}$. 
The green hatched band corresponds to $\phi=1$ and $\omega_{\rm c}=1$, which we used in our calculations
throughout this paper. The blue and red hatched bands are upper and lower limits set by reasonable choices
of the two parameters ($\phi$,~$\omega_{\rm c}$). 
In all cases we assumed a fixed accretion rate of $\dot{M}=\dot{M}_{\rm Edd}$. The location of the spin-up line
is simply shifted one order of magnitude in $\dot{P}$ down (up) for every order of magnitude $\dot{M}$
is decreased (increased).

It is important to realize that there is no universal spin-up line in the $P\dot{P}$--diagram. Only an upper limit. 
Any individual pulsar has its own spin-up line/location which in particular depends on its 
unknown accretion history ($\dot{M}$). Also notice that the dependence on the
magnetic inclination angle, $\alpha$ is much less pronounced when applying the Spitkovsky formalism for estimating
the B-field of the MSP compared to applying the vacuum dipole model. The difference in the location of spin-up lines
using $\alpha=90^{\circ}$ and $\alpha=0^{\circ}$ is only a factor of two in the Spitkovsky formalism. For a comparison, using the 
vacuum dipole model with a small magnetic inclination of $\alpha=10^{\circ}$ results in a spin-up line which 
is translated downwards by almost two orders of magnitude 
compared to its equivalent orthogonal rotator model, cf. the two red dashed lines in Fig.~\ref{fig:spinupline}.

If we assume that accretion onto the neutron star is indeed Eddington limited
then the three bands in Fig.~\ref{fig:spinupline} represent upper limits for the spin-up line for the 
given sets of ($\phi$,~$\omega _{\rm c}$).
Thus we can in principle use this plot to constrain ($\phi$,~$\omega _{\rm c}$) and hence the physics of disk--magnetosphere
interactions. 
The fully recycled pulsars B1937+21 and J0218+4232, and the mildly recycled pulsar PSR~J1841+0130, are interesting 
since they are located somewhat in the vicinity of the green spin-up line.
Any pulsar above the green line would imply that $\phi < 1$. 
The pulsar PSR~J1823$-$3021A is close to this limit. Usually the derived value of $\dot{P}$ for globular cluster
pulsars is influenced by the cluster potential. However, the $\dot{P}$ value for PSR~J1823$-$3021A 
was recently constrained from Fermi~LAT $\gamma$-ray observations \citep{faa+11} and for this reason we have included it here.
The high $\dot{P}$ values of the three Galactic field MSPs listed above: B1937+21, J0218+4232 and PSR~J1841+0130, 
imply that these MSPs are quite young. We discuss their true ages in Sections~\ref{subsubsec:twoPSRs} and \ref{subsubsec:1841}.

\subsubsection{Formation of a submillisecond pulsar}\label{subsubsec:submsMSP}
The possible existence of submillisecond pulsars ($P<1\,{\rm ms}$) is a long standing and intriguing question in the literature
since it is important for constraining the equation-of-state of neutron stars \citep[e.g.][]{hz89}. 
There has been intense, but so far unsuccessful, efforts to detect these objects in either radio or {X}-rays \citep{dam00,kjv+10,pat10b}. 
We note from equation~(\ref{eq:Peq0}) that if the inner edge of the accretion disk (roughly given by $r_{\rm mag}$) 
extends all the way down to the surface of the neutron star we obtain the smallest possible value for $P_{\rm eq}\sim 0.6\,{\rm ms}$
(depending on $M$ and assuming $\omega _{\rm c}=1$). It is possible to achieve $r_{\rm mag} \simeq R$ from a combination
of a large value of $\dot{M}$ and a small value of $B$, cf. equation~(\ref{eq:alfven}).

\subsection{Amount of mass needed to spin up a pulsar}\label{subsec:deltaM}

\subsubsection{Accretion-induced magnetic field decay}\label{subsubsec:Bdecay}
The widely accepted idea of accretion-induced magnetic field decay in neutron stars is based 
on observational evidence \citep[e.g.][]{tv86,smsn89,vb94}.
During the recycling process the surface B-field of pulsars is reduced by several orders of magnitude, 
from values of $10^{11-12}\,G$ to $10^{7-9}\,G$. However, it is still not understood if
this is caused by spin-down induced flux expulsion of the core proton fluxoids \citep{sbmt90},
or if the B-field is confined to the crustal regions and decays due to diffusion and Ohmic dissipation, 
as a result of a decreased electrical conductivity when heating effects set in from nuclear burning of the accreted material
\citep{gu94,kb97},
or if the field decay is simply caused by a diamagnetic screening by the accreted plasma 
-- see review by \citet{bha02} and references therein.
Although there has been attempts to model or empirically fit the magnetic field evolution of accreting neutron stars
\citep[e.g.][]{smsn89,zk06,wzz+11} the results are quite uncertain. One reason is that it is difficult to estimate
how much mass a given recycled pulsar has accreted. Furthermore, we cannot rule out the possibility
that some of these neutron stars may originate from the
accretion-induced collapse of a massive white dwarf, in which case they might be formed with a high B-field
near the end of the mass-transfer phase.\\
To model the B-field evolution for our purpose, which is to relate the spin period of pulsars
to the amount of mass accreted, we make the following assumptions:  
\begin{itemize} 
\item The B-field decays rapidly in the early phases of the accretion process via some 
      unspecified process (see above).
\item Accreting pulsars accumulate the majority of mass while spinning at/near equilibrium.
\item The magnetospheric boundary, $r_{\rm mag}$ is approximately
      kept at a fixed location for the majority of the spin-up phase,
      until the mass transfer ceases during the RLDP, 
      cf. Section~\ref{sec:RLDP}. 
\item During the RLDP the B-field of an AXMSP can be considered to be constant 
      since very little envelope material ($\sim 0.01\,M_{\odot}$) remains to be transfered from its donor at this stage.
\end{itemize} 

\subsubsection{Accreted mass vs final spin period relation}\label{subsubsec:PspinDeltaM}
The amount of spin angular momentum added to an accreting pulsar is given by: 
\begin{equation}
  \Delta J_\star = \int n(\omega,t)\,\dot{M}(t)\,\sqrt{GM(t)r_{\rm mag}(t)}\,\xi (t)\;dt
  \label{eq:Jacc}
\end{equation}
where $n(\omega)$ is a dimensionless torque, see Section~\ref{sec:RLDP} for a discussion.
Assuming $n(\omega)=1$, and $M(t)$, $r_{\rm mag}(t)$ and $\xi(t)$ to be roughly constant during the major part of the spin-up phase
we can rewrite the expression and obtain a simple formula (see also equation~\ref{eq:deltaJ}) for the amount of matter needed to spin up
the pulsar:
\begin{equation}
  \Delta M \simeq \frac{2\pi I}{P\sqrt{GMr_{\rm mag}}\,\xi}
  \label{eq:deltaM}
\end{equation}
Note, that the initial spin angular momentum of the pulsar prior to accretion is negligible given that
$\Omega _0 \ll \Omega _{\rm eq}$.  
To include all numerical scaling factors properly we can insert 
equations~(\ref{eq:alfven}),~(\ref{eq:Bspitkovsky}) and ~(\ref{eq:spinupline}) into equation~(\ref{eq:deltaM}), 
recalling that $r_{\rm mag}=\phi\cdot r_{\rm A}$, and we find: 
\begin{equation} 
     \Delta M_{\rm eq} = I \left( \frac{\Omega_{\rm eq}^4}{G^2M^2} \right) ^{1/3} \,f(\alpha,\xi,\phi,\omega_c)
  \label{eq:deltaMfinal}
\end{equation}
where $f(\alpha,\xi,\phi,\omega_c)$ is some dimensionless number of order unity.
Once again we can apply the relation between moment of inertia and mass of the neutron star 
(see Section~\ref{subsubsec:eddlimit}) and we obtain
a simple convenient expression to relate the amount of mass to be accreted in order to spin~up
a pulsar to a given (equilibrium) rotational period:
\begin{equation} 
     \Delta M_{\rm eq} = 0.22\,M_{\odot}\; \frac{(M/M_{\odot})^{1/3}}{P_{\rm ms}^{4/3}}
  \label{eq:deltaMfinalfit}
\end{equation}
assuming that the numerical factor $f(\alpha,\xi,\phi,\omega_c)=1$.\\
In the above derivation we have neglected minor effects related to release of gravitational binding energy of the accreted material
-- see e.g. equation~(22) in \citet{ts99}, and in particular \citet{bzhf11} and \citet{bag11} for a more detailed, general discussion
including various equations of state, general relativity and the critical mass shedding spin limit. 
However, since the exchange of angular momentum takes place near the magnetospheric boundary
the expression in equation~(\ref{eq:deltaMfinalfit}) refers to the baryonic mass accreted from 
the donor star. To calculate the increase in (gravitational) mass of the pulsar one must
apply a reducing correction factor of $\sim \!0.85-0.90$, depending on the 
neutron star equation-of-state \citep{ly89}.

In Fig.~\ref{fig:deltaM} we show the amount of mass, $\Delta M_{\rm eq}$ needed to spin~up a pulsar to a given spin period.
The value of $\Delta M_{\rm eq}$ is a strongly decreasing function of the pulsar spin period, $P_{\rm eq}$.
For example, considering a pulsar with a final mass of $1.4\,M_{\odot}$ and a recycled spin period
of either 2~ms, 5~ms, 10~ms or 50~ms requires accretion of 
$0.10\,M_{\odot}$, $0.03\,M_{\odot}$, $0.01\,M_{\odot}$ or $10^{-3}\,M_{\odot}$, respectively.
Therefore, it is no surprise that observed recycled pulsars with massive companions ({CO}~WD, {O}-{Ne}-{Mg}~WD or NS)
in general are much more slow rotators -- compared to BMSPs with {He}~WD companions --
since the progenitor of their massive companions evolved on a relatively short timescale,
only allowing for very little mass to be accreted by the pulsar, cf.~Section~\ref{sec:obs-spin}.

\subsubsection{Comparison with other work}\label{subsubsec:comparisonM}
\begin{figure}
\begin{center}
  \includegraphics[width=0.35\textwidth, angle=-90]{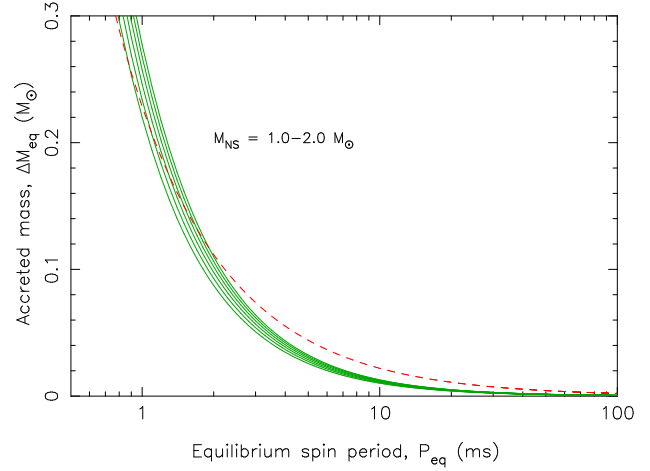}
  \caption[]{The amount of mass needed to spin up a pulsar as a function of its equilibrium spin period using equation~(\ref{eq:deltaMfinalfit}).
             The different green curves correspond to various neutron star masses in steps of $0.2\,M_{\odot}$, increasing upwards.
             The dashed red curve is the expression from equation~(\ref{eq:Pmin}) -- see text. 
    }
\label{fig:deltaM}
\end{center}
\end{figure}

The simple expression in equation~(\ref{eq:deltaMfinalfit}) was also found by \citet{acrs82}. 
Alternatively, one can integrate equation~(\ref{eq:deltaM})
directly which yields the maximum spin rate (minimum period)
that can be attained by a neutron star accelerated from rest \citep{lp84}:
\begin{equation} 
      P_{\rm min} = \frac{3\pi\,I}{\sqrt{G\,r_{\rm mag}}\,\xi}\,\left( M^{3/2}-M_{\rm init}^{3/2}\right) ^{-1} 
  \label{eq:Pmin}
\end{equation}
where $M_{\rm init}=M-\Delta M_{\rm eq}$ is the mass of the neutron star prior to accretion.
In this expression it is still assumed that $I$ and $r_{\rm mag}$ remain constant during the accretion phase.
For comparison, the expression above
is also plotted in Fig.~\ref{fig:deltaM} as a dashed line assuming $\Delta M_{\rm eq} \ll M$, $\xi=1$ and $r_{\rm mag}=22\,{\rm km}$
(for example, if $B=10^8\,{\rm G}, \dot{M}=0.1\,\dot{M}_{\rm Edd}, M=1.4\,M_{\odot}$, $\phi=1$, see equation~(\ref{eq:alfven})). 
Although the latter expression is simplified with a chosen numerical value of $r_{\rm mag}$, and 
the two expressions differ by more than a factor of two if $P>10\,{\rm ms}$,
the overall match is fairly good. 

One the one hand, the value of $\Delta M_{\rm eq}$ should be regarded as a lower limit to the actual amount of material 
required to be transfered to the neutron star, even at sub-Eddington rates, 
since a non-negligible amount may be ejected from the pulsar magnetosphere due to magnetodipole wave pressure
or the propeller effect \citep{is75}. Furthermore, accretion disk instabilities \citep{pri81,vpa96} 
are also responsible for ejecting part of the transfered material. Recently it was demonstrated by
\citet{akk+12} that the accretion efficiency in some cases is less than 50~per~cent, even in short orbital period
binaries accreting at sub-Eddington levels.
On the other hand, we did not take into account the possibility of a more efficient angular momentum transfer
early in the accretion phase where the value of $r_{\rm mag}$ (the lever arm of the torque) could have been
larger if the B-field did not decay rapidly. In this respect it is somewhat surprising that \citet{wzz+11}, 
using a model for accretion induced B-field decay, find that it requires more mass to be accreted 
to obtain a given spin period (see their fig.~6) compared to our work. This could perhaps (partly) be
related to the fact that $\Omega$ is small during the initial accretion phases.

\subsection{Spin-relaxation timescale}\label{subsec:timescale}
Above we have calculated the amount of mass needed to spin up a pulsar to a given equilibrium spin period.
However, we must be sure that the accretion-torque can actually transmit this acceleration on a timescale 
shorter than the mass-transfer timescale. To estimate the spin-relaxation timescale (the time needed to spin up a
slowly-rotating neutron star to spin equilibrium) one can simply consider: $t_{\rm torque}=J/N$ 
where $J=2\pi I/P_{\rm eq}$ and $N=\dot{M}\sqrt{GMr_{\rm mag}}\,\xi$ which yields:
\begin{eqnarray} 
\label{eq:timescale}
    t_{\rm torque} & = & I\,\left(\frac{4G^2M^2}{B^8R^{24}\dot{M}^3}\right) ^{1/7} \frac{\omega _c}{\phi^2\,\xi} \\ \nonumber
           & \simeq & 50\,{\rm Myr}\quad B_8^{-8/7}\left(\frac{\dot{M}}{0.1\,\dot{M}_{\rm Edd}}\right) ^{-3/7} \left(\frac{M}{1.4\,M_{\odot}}\right) ^{17/7}
\end{eqnarray} 
or equivalently, $t_{\rm torque}=\Delta M/\dot{M}$, see equation~(\ref{eq:deltaM}). 
If the duration of the mass-transfer phase, $t_X$ is shorter than $t_{\rm torque}$ (this is the case if the mass-transfer rate 
$|\dot{M}_2|\gg\dot{M}$) then the pulsar will not be fully recycled.
In the following section we examine the importance of $t_{\rm torque}$ relative to the RLO turn-off timescale
at the end of the mass-transfer process and how their ratio may affect the pulsar spin period.  

%%%%%%%%%%%%%%%%%%%%%%%%%%%%%%%%%%%%%%%%%%%%%%%%%%%%%%%%%%%%%%%%%%%%%%%%%%%%%%%%

\section{Roche-lobe decoupling phase -- RLDP} \label{sec:RLDP}
An important effect on the final spin evolution of accreting pulsars, related to the 
Roche-lobe decoupling phase (RLDP), has recently been demonstrated \citep{tau12}. The rapidly decreasing mass-transfer rate during the RLDP
results in an expanding magnetosphere which may cause a significant braking torque to act on the spinning pulsar. 
It was shown that this effect can explain why the recycled radio MSPs (observed {\it after} the RLDP) 
are significantly slower rotators compared to 
the rapidly spinning accreting {X}-ray MSPs (observed {\it before} the RLDP).
This difference in spin periods was first noted by \cite{hes08}. However, 
only for MSPs with $B> 10^8\,{\rm G}$ can the difference in spin periods be partly understood from regular magneto-dipole 
and plasma current spin-down of the recycled radio MSPs.
In this section we investigate the RLDP~effect on IMXBs and LMXBs
and demonstrate different outcomes for BMSPs with {CO} and {He}~WD companions.
The purpose of the computations is to follow the  
spin evolution of the accreting {X}-ray millisecond pulsar (AXMSP) when the donor star decouples from its Roche-lobe, and to calculate the
initial spin period of the recycled radio pulsar once the mass-transfer ceases. 
We model the main effect (the RLDP which causes the magnetosphere to expand dramatically) on the general spin evolution  
and ignore other, less known or somewhat hypothetical, dynamical effects during this epoch, such as 
warped disks, disk-magnetosphere instabilities, variations in the mass transfer rate caused by {X}-ray irradiation effects of the donor star,  
or transitions between a Keplerian thin disk
and a sub-Keplerian, advection-dominated accretion flow (ADAF) \citep[e.g.][and references therein]{st93,nbc+97,ywv97,vcpw98,lw98,lm04,dl06,cpfw+11}.
Although many of these effects may be less important for the RLDP they could perhaps explain
some of the frequent torque reversals observed in {X}-ray pulsars \citep{bcc+97}.

\subsection{Onset of a propeller phase}\label{subsec:propeller}
In the final stages of the {X}-ray phase, when the donor star is just about to detach from its Roche-lobe, the mass-transfer
rate decreases significantly. This causes the ram pressure of the incoming flow to decrease whereby the 
magnetospheric boundary, and thus the coupling radius $r_{\rm mag}$, moves outward. 
When this boundary moves further out than the corotation radius ($r_{\rm mag} > r_{\rm co}$) given by:
\begin{eqnarray} 
\label{eq:r_co}
    r_{\rm co} & = & \left( \frac{GM}{\Omega _\star ^2}\right) ^{1/3} \\ \nonumber 
           & \simeq & 17\,{\rm km}\;\; P_{\rm ms}^{2/3}\left(\frac{M}{1.4\,M_{\odot}}\right) ^{1/3}
\end{eqnarray} 
a centrifugal barrier arises since the plasma flowing towards the neutron star couples to the field lines in a
super-Keplerian orbit. The material is therefore presumably ejected away from the neutron star in this propeller phase \citep{is75}.
This ejection of material causes exchange of angular momentum between the now relatively fast spinning neutron star and the slower 
spinning material at the edge of the inner disk. The result is a braking torque which acts to slow down the spin of the pulsar.
This causes the corotation radius to move outwards too and the subsequent evolution is determined by the rate at which these two radii 
expand relative to one another. However, during the RLDP $\dot{M}$ decreases too rapidly for $r_{\rm co}$ to keep up with the
rapidly expanding $r_{\rm mag}$ and the equilibrium spin phase comes to an end.

In our model we assumed, in effect, that the inner edge of the accretion disk, $r_{\rm disk}$ follows $r_{\rm mag}$ and that the centrifugally expelled 
material has sufficient kinetic energy to be gravitationally unbound during the propeller phase when $r_{\rm mag}>r_{\rm co}$.
These assumptions may not be entirely correct \citep{st93,rfs04}. 
For example, \citet{ds10,ds11,ds12} recently demonstrated that accretion disks may be trapped near the corotation radius, even at low accretion rates.
In this state the accretion can either continue to be steady or become cyclic. Furthermore, based on energy considerations they point out that
most of the gas will not be ejected if the Keplerian corotation speed is less than the local escape speed. 
This will be the case if 
$r_{\rm disk}<1.26\,r_{\rm co}$.
However, this trapped state only occurs in a narrow region around the corotation radius ($\Delta r \ll r_{\rm co}$) 
and given that $r_{\rm mag}/r_{\rm co}$ often reaches a factor of $3-5$ in our models at the end of the RLDP 
we find it convincing that the RLDP effect of braking the spin rates of the recycled pulsars is indeed present
(see \citet{tau12} for further details).

\subsubsection{The magnetosphere--disk interaction}\label{subsubsec:magdisk}
In our numerical calculations of the propeller phase we included the effect of additional spin-down torques, acting on the neutron star,
due to both magnetic field drag on the accretion disk \citep{rfs04} as well as magnetic dipole radiation (see equation~(\ref{eq:dipole})),
although these effects are usually not dominant. 
It should be noted that the magnetic stress in the disk is also related to the critical fastness parameter, $\omega _c$ \citep{gl79b}. 
We follow \citet{tau12} and write the total spin torque as:
\begin{equation} 
\label{eq:drag}
  N_{\rm total} = n(\omega)\left(\dot{M}\sqrt{GMr_{\rm mag}}\,\xi + \frac{\mu ^2}{9r_{\rm mag}^3}\right) - \frac{\dot{E}_{\rm dipole}}{\Omega}
\end{equation} 
where
\begin{equation} 
\label{eq:dimlesstorque}
  n(\omega)=\tanh \left( \frac{1-\omega}{\delta _\omega} \right)
\end{equation} 
is a dimensionless function, depending on the fastness parameter, 
$\omega = \Omega _\star /\Omega_{\rm K}(r_{\rm mag}) = (r_{\rm mag}/r_{\rm co})^{3/2}$, which is introduced
to model a gradual torque change in a transition zone near the magnetospheric boundary.
The width of this zone has been shown to be small \citep{st93}, corresponding to $\delta _\omega \ll 1$ and 
a step-function-like behavior $n(\omega)=\pm 1$. 
In our calculations presented here we used $\delta _\omega =0.002$, $\xi=1$, $r_{\rm disk}=r_{\rm mag}$  
and also assumed the moment of inertia, $I$, to be constant during the RLDP. 
The latter is a good approximation since very little material is accreted during this terminal stage of the RLO.

\subsubsection{Radio ejection phase}\label{subsubsec:radioeject}
As the mass-transfer rate continues to decrease, 
the magnetospheric boundary eventually crosses the light-cylinder radius, $r_{\rm lc}$ of the neutron star given by:
\begin{equation} 
\label{eq:r_lc}
    r_{\rm lc} = c/\Omega_\star \simeq 48\,{\rm km}\;\; P_{\rm ms}
\end{equation} 
When $r_{\rm mag} > r_{\rm lc}$ 
the plasma wind of the pulsar can stream out along the open field lines, providing the necessary condition for 
the radio emission mechanism to turn on \citep[e.g.][]{mic91,spi08}. %% See also Hankins, Rankin & Eilek (2009) white paper
Once the radio millisecond pulsar is activated the presence of the plasma wind may prevent further accretion from the now weak flow of material
at low $\dot{M}$ \citep{krst88}. As discussed by \citet{bpd+01,bdb02} this will be the case 
when the total spin-down pressure of the pulsar (from magneto-dipole radiation and the plasma wind) exceeds the inward pressure
of the material from the donor star, i.e. if: 
$\dot{E}_{\rm rot}/(4\pi r^2\,c) > P_{\rm disk}$.
However, it is interesting to notice that the recycled pulsar PSR~J1023+0038 \citep{asr+09} recently (within a decade) 
turned on its radio emission 
and it would indeed be quite a coincidence if we have been lucky enough to catch that moment -- unless
recycled pulsars evolve through a final phase with recurrent changes between accretion and radio emission modes.   

\subsection{Slow RLDP}\label{subsec:slowRLDP}
\begin{figure*}
\begin{center}
  \includegraphics[width=0.96\textwidth, angle=0]{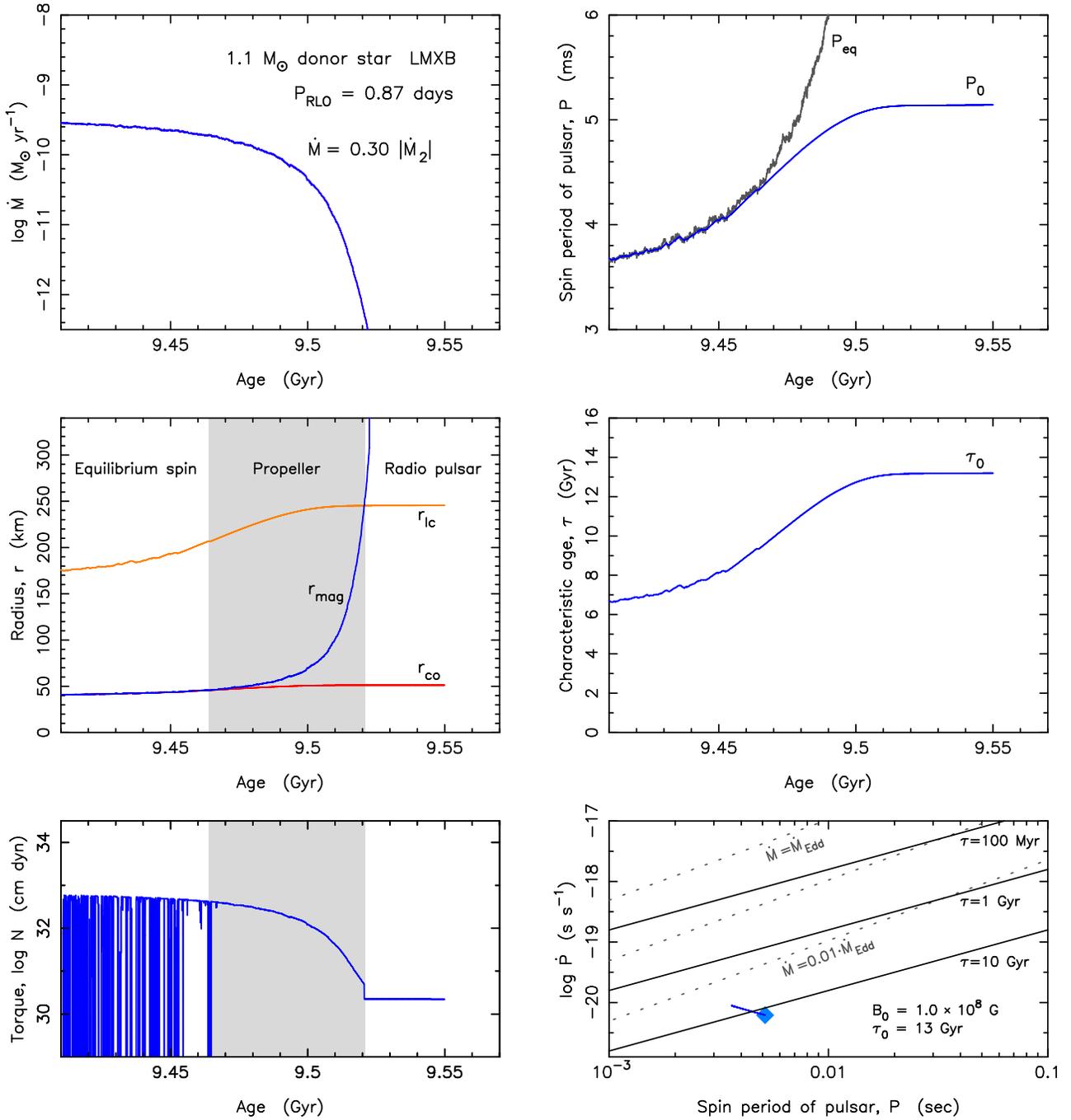}
  \caption[]{
    Example of a slow Roche-lobe decoupling phase (RLDP) and the final spin evolution of a pulsar formed with a {He}~WD companion in an LMXB.
    The left side panels show: The accretion rate of the pulsar, $\dot{M}$
    as a function of the total age of the donor star (top); the evolution of the location of the magnetospheric boundary, $r_{\rm mag}$,
    the corotation radius, $r_{\rm co}$ and the light-cylinder radius, $r_{\rm lc}$ (centre);
    and the resulting braking torque (bottom). The right side panels show: The evolution of the pulsar spin period as a function 
    of the total age of the donor star (top); the evolution of the pseudo characteristic age of the pulsar (centre); 
    and the RLDP evolutionary track leading to the birth location 
    in the $P\dot{P}$--diagram shown as a blue diamond (bottom). 
    The three epochs of the RLDP: {\it equilibrium spin}, {\it propeller phase} and {\it radio pulsar} are easily identified in the left side panels.
    The propeller phase is grey shaded. In these model calculations we assumed $\sin \alpha = \phi = \xi = \omega_{\rm c}=1$ and $\delta _\omega = 0.002$.
    It is interesting to notice how the changes in the accretion rate, $\dot{M}$ affects the magnetospheric coupling radius, $r_{\rm mag}$ which
    again affects the spin-down torque and the equilibrium spin period, cf. equations~(\ref{eq:alfven}), (\ref{eq:deltaJ}) and
    (\ref{eq:Peq0}). 
    The reason why the spin of the pulsar (blue line, upper right panel) decouples from its equilibrium spin (black line) is that the
    spin-down torque cannot transmit the deceleration on a timescale short enough for the pulsar to adapt to its new equilibrium.
    In this example, the duration of the RLDP is a significant fraction of the spin-relaxation timescale of the accreting neutron star
    and thus the RLDP has a momentous impact on the rotational evolution. 
    The birth spin period, $P_0$ of this recycled pulsar is seen to deviate substantially from its original $P_{\rm eq}$ calculated at the onset of the RLDP. 
    Furthermore, the effect on the
    characteristic age for this BMSP is also quite significant. During the 56~Myr of the RLDP it doubles to become
    $\tau =13\,{\rm Gyr}$ at {\it birth}. Hence, many MSPs which appear to be old according to their $\tau$
    may in fact be quite young. See text for further discussions.
    }
\label{fig:spinLMXB}
\end{center}
\end{figure*}

\begin{figure*}
\begin{center}
  \includegraphics[width=0.96\textwidth, angle=0]{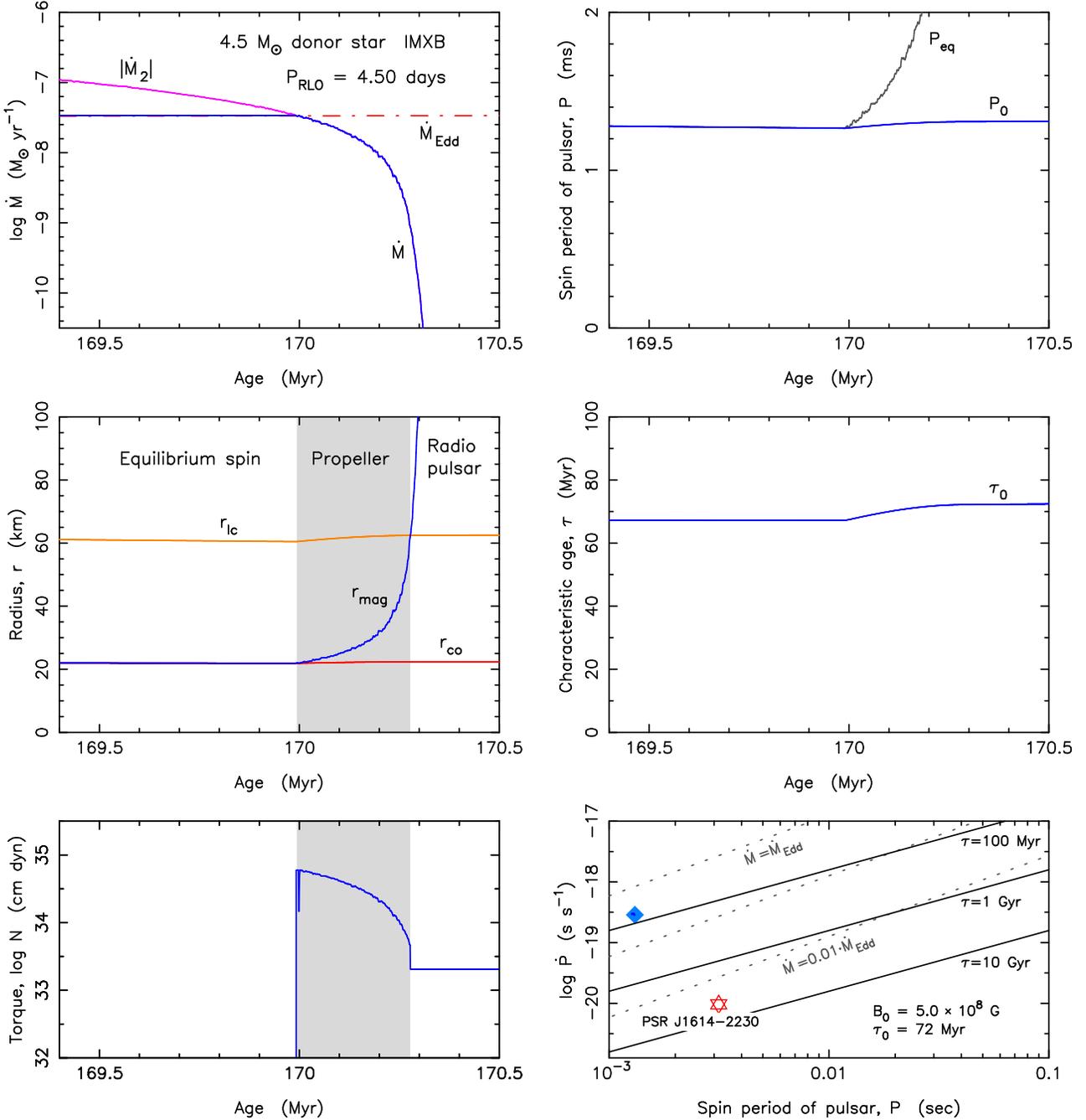}
  \caption[]{
    Example of a rapid Roche-lobe decoupling phase (RLDP) in an IMXB using our mass-transfer progenitor model of \psr. 
    The spin evolution of the pulsar is shown during the RLDP when the IMXB donor star terminated its mass transfer.
    For a description of the various panels and their labels, see Fig.~\ref{fig:spinLMXB}.
    In the $P\dot{P}$--diagram in the lower right panel the birth location, immediately following the RLDP, of 
    the recycled radio pulsar is shown as a blue diamond for $\sin \alpha = \phi = \xi = \omega _c =1$. 
    The red star indicates the presently observed location of \psr.
    In Section~\ref{subsec:ppdot-evol} we argue that \psr\ is more likely to have evolved with $\sin \alpha < 1$, $\phi>1$ and $\omega _c <1$
    which results in a birth location much closer to its present location in the $P\dot{P}$--diagram. 
    In IMXBs where the donor star evolves on a relatively short timescale,
    the RLDP is rapid. Therefore, the spin evolution of the accreting {X}-ray pulsar (see blue line in upper right panel) freezes up near the original value of
    $P_{\rm eq}$, i.e. the recycled radio pulsar is born with a spin period, $P_0$ which is close to the equilibrium spin period of
    the accreting {X}-ray pulsar calculated at the time when the donor star begins to decouple from its Roche-lobe. 
    In this IMXB model the mass-transfer rate, $\dot{M}>\dot{M}_{\rm Edd}$ and the accretion onto the neutron star is therefore limited by the Eddington 
    mass-accretion rate (red dot-dashed line in the upper left panel).
    This explains the why the spin of the pulsar, $P=P_{\rm eq}\simeq const.$ (top right panel) until $\dot{M}<\dot{M}_{\rm Edd}$, 
    as opposed to the situation in Fig.~\ref{fig:spinLMXB}.
    }
\label{fig:spinIMXB}
\end{center}
\end{figure*}

In Fig.~\ref{fig:spinLMXB} we have shown a model calculation of the RLDP 
of an LMXB with an original donor star mass of $1.1\,M_{\odot}$. The outcome was the 
formation of a BMSP with $P=5.2\,{\rm ms}$ (assuming $B=1.0\times 10^8\,{\rm G}$) and a {He}~WD companion of mass $0.24\,M_{\odot}$, orbiting with a period of 5.0~days.
It has been argued by \citet{ts99,akk+12} that the majority of the transfered material in some LMXBs is 
lost from the system, even for accretion at sub-Eddington rates.
Therefore, we assumed an effective accretion efficiency of 30~per~cent in our LMXB model, i.e.
$\dot{M}=0.30\,|\dot{M}_2|$, where $|\dot{M}_2|$ is the RLO mass-transfer rate from the donor star.
The accreted\footnote{Strictly speaking this is the estimated mass-transfer rate received from the inner edge of the accretion disk.
Some of this material is ejected during the propeller phase.}
mass-transfer rate $\dot{M}$ is shown in the upper left panel.\\  
The three phases:
{\it equilibrium spin} (corresponding to $r_{\rm mag} \simeq r_{\rm co}$), {\it propeller phase} ($r_{\rm co} < r_{\rm mag} < r_{\rm lc}$)
and {\it radio pulsar} ($r_{\rm mag} > r_{\rm lc}$) are clearly identified in this figure. 
During the equilibrium spin phase the rapidly alternating sign changes of the torque 
(partly unresolved on the graph in the lower left panel) reflect 
small oscillations around the semi-stable equilibrium, corresponding to successive small episodes of spin~up and spin~down.
The reason is that the relative location of $r_{\rm mag}$ and $r_{\rm co}$
depends on the small fluctuations in $\dot{M}$. Despite applying an
implicit coupling scheme in the code and ensuring that our time steps
during the RLDP (about $8\times 10^4\,{\rm yr}$) are much smaller than the
duration of the RLDP ($\sim \! 10^8\;{\rm yr}$), our calculated
mass-transfer rates are subject to minor numerical oscillations. However,
these oscillations could in principle be physical within the frame of our
simple model. Examples of physical perturbations that could cause real
fluctuations in $\dot{M}$, but on a much shorter timescale, are accretion
disk instabilities and clumps in the transfered material from the donor
star (our donor stars have active, convective envelopes during the RLDP).  

At some point the equilibrium spin is broken. Initially, the spin can remain in equilibrium by adapting to the decreasing
value of $\dot{M}$. However, further into the RLDP the result is that $r_{\rm mag}$ increases on a timescale faster than
the spin-relaxation timescale, $t_{\rm torque}$ at which the torque can transmit the effect of deceleration to
the neutron star, and therefore $r_{\rm mag} > r_{\rm co}$ 
\citep[see][and supporting online material therein]{tau12}.
During the propeller phase the resultant accretion torque acting on the neutron star is negative, i.e. a spin-down torque. When the radio
pulsar is activated and the accretion has come to an end, the spin-down torque is simply caused by the magneto-dipole radiation combined 
with the pulsar wind (see Section~\ref{subsubsec:B}).

The aim here is to calculate the initial spin period of the recycled radio pulsar, $P_0$ once the mass-transfer ceases. 
In this LMXB model calculation the value of the spin period increased significantly during the RLDP.
The reason for this is that the duration of the propeller phase (i.e. the RLDP) is a substantial fraction of the
spin-relaxation timescale, $t_{\rm torque}$. Using equation~(\ref{eq:timescale}) 
we find $t_{\rm torque}=195\,{\rm Myr}$ whereas the RLDP
lasts for $t_{\rm RLDP}\simeq 56\,{\rm Myr}$ 
which is a significant fraction of $t_{\rm torque}$ ($t_{\rm RLDP}/t_{\rm torque}=0.29$). Therefore, the RLDP has quite a significant
impact on the spin evolution of the neutron star. It is seen that the spin period increases from 3.7~ms to 5.2~ms 
during this short time interval, i.e. the pulsar loses 50~per~cent of its rotational energy during the RLDP. 
As shown by \citet{tau12} this RLDP~effect is important for understanding the apparent difference in spin period distributions between 
AXMSPs\footnote{The majority of the AXMSPs have very small $P_{\rm orb}$ (typically 1--2 hours)
and substellar companions ($<0.08\,M_{\odot}$) and hence most of these AXMSPs are not true progenitors of the general population of
radio MSPs. However, the important thing is that the pulsars are spun up to rotational periods which seem to be almost independent
of orbital period up to 200~days, as seen in the central panel of Fig.~\ref{fig:corbet3}.} 
and radio MSPs \citep{hes08}. 
The black line labelled ``$P_{\rm eq}$'' in the upper right panel reflects the evolution of $P$ in case the neutron star
was able to instantly re-adjust itself to the changing $P_{\rm eq}$ (equation~\ref{eq:Peq0}) during the RLDP.
However, the neutron star possesses a large amount of rotational inertia and such rapid changes in $P$ are not
possible to transmit to the neutron star given the limited torque acting on it.
Therefore the spin evolution terminates at the end of the blue line labelled ``$P_0$''.
The small fluctuations of $P_{\rm eq}$ during the equilibrium spin phase reflect fluctuations in $\dot{M}(t)$.

The RLDP~effect
also causes a large impact on the characteristic age, $\tau _0$, of the radio pulsar at birth. In the central right panel  
we show how $\tau _0$ doubles in value from 6.7~Gyr to 13~Gyr during the RLDP (see also the solid line ending at the blue diamond in 
the $P\dot{P}$--diagram in the lower right panel). 
The characteristic age at birth is given by: 
$\tau _0 \equiv P_0/2\dot{P}_0$, where $\dot{P}_0$ can be estimated from the assumed B-field of the pulsar at this epoch.
Actually, the ``characteristic age'' of the pulsar during the RLDP (which is plotted in the central right panel) is a pseudo age which is only relevant at
the end point ($\tau _0$) when the accretion has stopped. 
In order to construct the track leading to $\tau _0$ we assumed $B^2 \propto P\dot{P}=const.$ from equation~(\ref{eq:B}),
leading to $\tau \propto P^2$. 
During the RLDP it is a good approximation to assume
a (final) constant value of the neutron star surface magnetic flux density, $B$ 
since 99~per~cent of the accretion onto the neutron star occurs before the final termination stage of the RLDP. 
In this case we assumed a final B-field strength of $B=1\times 10^8\,{\rm G}$ -- a typical value for BMSPs with {He}~WD companions.\\
This example clearly reflects why characteristic ages of MSPs are untrustworthy as true age indicators
since here the MSP is born with $\tau _0 \simeq 13\,{\rm Gyr}$ (see also \citet{tau12}).
This fact is important since 
the characteristic ages of recycled pulsars are often compared to the cooling ages of 
their white dwarf companions \citep[e.g. see discussion in][]{vbjj05}. 

\subsection{Rapid RLDP ($P_{\rm eq}$ freeze-up)}\label{subsec:rapidRLDP}
In Fig.~\ref{fig:spinIMXB} we have plotted the RLDP during the final stages of the IMXB phase~AB
using our Case~A model calculation of \psr\ presented in Paper~I. The initial conditions were a $4.5\,M_{\odot}$ donor star in an IMXB
with an orbital period of 2.20~days. The final system was a BMSP with a $0.50\,M_{\odot}$ {CO}~WD orbiting a $1.99\,M_{\odot}$ recycled pulsar
with an orbital period of 8.7~days -- resembling the characteristics of the \psr\ system.
The accretion rate onto the neutron star, $\dot{M}$ (see blue line in upper left panel)
is Eddington limited (red dot-dashed line) and hence $\dot{M}=\min\left( |\dot{M}_2|, \dot{M}_{\rm Edd}\right)$
where $|\dot{M}_2|$ represents the mass-transfer rate from the donor star (purple line).
When the age of the donor star is $\sim\!170.0$~Myr the mass-transfer rate becomes sub-Eddington and $\dot{M}$ decreases with $|\dot{M}_2|$.
From this point onwards $r_{\rm mag}$ increases as expected, as a result of the decreasing ram pressure, and it increases at a rate where $r_{\rm co}$ cannot keep up with it
(see central panel, left column) and therefore the pulsar enters the propeller phase. After less than 0.3~Myr 
the accretion phase, and thus the propeller phase,
is terminated when $r_{\rm mag} > r_{\rm lc}$ and the radio ejection mechanism is activated. 

During the equilibrium spin phase the net accretion torque acting on the pulsar is close to zero. This is not only due to our transition zone
approximation (equation~(\ref{eq:dimlesstorque})) where the torque vanishes when $r_{\rm mag}\simeq r_{\rm co}$. If the transition zone
was arbitrary thin ($\delta _\omega \lll 1$), and $n(\omega)$ would be a step-like function going from $+1$ to $-1$ (i.e. switch mode), there would be
rapidly alternating sign changes of the torque during the equilibrium spin phase. Whereas small fluctuations in $\dot{M}$
would lead to such rapid torque oscillations for sub-Eddington accretion (as seen in Fig.~\ref{fig:spinLMXB}) the sharply defined Eddington limit
applied here prevents such fluctuations during the equilibrium spin phase. However, the time-averaged torque would be close to zero
in any case. 

In this IMXB model calculation the value of the spin period only increased very little during the RLDP.
The reason for this is that the duration of the RLDP ($t_{\rm RLDP}\simeq0.28\,{\rm Myr}$) is relatively short compared to the
spin-relaxation timescale ($t_{\rm torque}=6.8\,{\rm Myr}$). 
The small ratio $t_{\rm RLDP}/t_{\rm torque}\simeq 0.04$ at the onset of the RLDP 
causes the spin period to ``freeze'' at the original value of $P_{\rm eq}$ \citep[see also][]{rst89}. 
Note, the values quoted above as well as rough values of
$t_{\rm torque}=(2\pi/P)\,I/N$ and $r_{\rm mag}\simeq r_{\rm co}\simeq 22\,{\rm km}$ can be checked by reading numbers from the plots and  
using equations~(\ref{eq:alfven}) and (\ref{eq:r_co}). The relevant numbers are:
$B=5\times 10^8\,{\rm G}$ (a typical value for BMSPs with {CO}~WD companions), 
$\dot{M}\simeq\dot{M}_{\rm Edd}$, $M\simeq 1.99\,M_{\odot}$, $P_{\rm eq}=1.28\,{\rm ms}$ and $\log N=34.8$.
For the spin-up process discussed here we assumed again $\sin \alpha = \phi= \xi = \omega _c =1$.

\subsubsection{Rapid RLDP with a helium star donor}\label{subsubsec:rapidRLDPHe}
As discussed in Section~\ref{sec:CEspin} (and seen in Fig.~\ref{fig:HeNS}) helium donor stars evolve on a short nuclear timescale compared to
a normal hydrogen-rich donor. This leads to a rapid accretion turn-off lasting only a few $10^4\,{\rm yr}$
and therefore $t_{\rm RLDP}/t_{\rm torque} \ll 1$, which also in these systems causes 
$P_0$ to ``freeze'' at the original value of $P_{\rm eq}$. 

\subsection{RLDP in IMXB vs LMXB systems}\label{subsec:RLDPvs}
The major difference between 
the RLDP in an IMXB (often leading to a BMSP with a {CO}~WD) and the RLDP in an LMXB (leading to a BMSP with a {He}~WD) is the 
time duration of the RLDP relative to the spin-relaxation timescale, i.e. the $t_{\rm RLDP}/t_{\rm torque}$-ratio. 
The mass-transfer in {X}-ray binaries proceeds much faster in IMXBs compared to LMXBs. Hence, also the
termination of the RLDP is shorter in IMXBs, and therefore the $t_{\rm RLDP}/t_{\rm torque}$-ratio is smaller, compared to LMXBs. 
The conclusion is that, in general, we expect a significant RLDP~effect only in LMXBs. 
(It may be possible, however, that the RLDP~effect could play a more significant role in some IMXB systems 
if they leave behind a neutron star with a relatively high B-field, thereby decreasing $t_{\rm torque}$).

%%%%%%%%%%%%%%%%%%%%%%%%%%%%%%%%%%%%%%%%%%%%%%%%%%%%%%%%%%%%%%%%%%%%%%%%%%%%%%%%

\section{Observed spin-period distributions}\label{sec:obs-spin}
We now investigate if the empirical pulsar spin data can be understood in view of the theoretical modelling
presented in the last couple of sections. 
\begin{figure}
\begin{center}
  \includegraphics[width=0.35\textwidth, angle=-90]{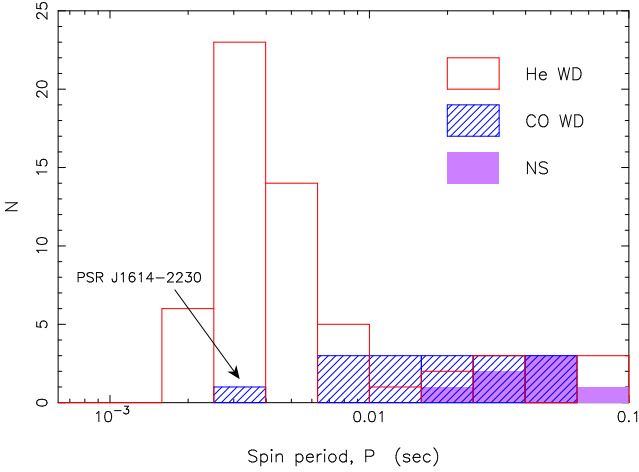}
  \caption[]{The observed spin period distribution of recycled pulsars in the Galactic disk with $P<100\,{\rm ms}$ for pulsars with 
             {He}~WD and {CO}~WD companions. The median spin periods for these populations are 3.9~ms and 16~ms, respectively.
             For comparison we have also shown the spin distribution of those recycled pulsars with neutron star companions
             and $P<100~{\rm ms}$.
    }
\label{fig:histo}
\end{center}
\end{figure}

\subsection{Recycled pulsars with WD companions}\label{subsec:NSWD}
In Fig.~\ref{fig:histo} we have plotted histograms of the observed spin period distribution of recycled pulsars
with {He}~WD and {CO}~WD companions. There are 57 known recycled pulsars in the Galactic disk with a {He}~WD companion
and a spin period $P<100\,{\rm ms}$ (with or without a measured $\dot{P}$) and these pulsars have a median spin period of 3.9~ms 
(the average spin period is 10~ms).
The median spin period of the 16 known recycled pulsars with a {CO}~WD companion and $P<100\,{\rm ms}$, however, is 16~ms 
(the average value is 23~ms). 
All observed BMSPs have lost rotational energy due to magneto-dipole radiation and a pulsar wind since they
appeared as recycled pulsars. Hence, their current spin period, $P$ is larger than their $P_{\rm eq}$.
Unfortunately, we have no empirical constraints on the spin-down torque acting on recycled radio MSPs
and therefore their braking index remains unknown, cf. Section~\ref{sec:trueages}.
Furthermore, in Section~\ref{sec:RLDP} we demonstrated that BMSPs formed in LMXBs lose rotational energy during the RLDP.
Correcting for both of these spin-down effects we here simply assume $P=\sqrt{2}\,P_{\rm eq}$. 
The resulting median values of $P_{\rm eq}$ 
are thus found to be 2.8~ms and 11~ms, respectively.
For a $1.4\,M_{\odot}$ neutron star these median equilibrium spin periods correspond to
$\Delta M_{\rm eq}=0.06\,M_{\odot}$ and $\Delta M_{\rm eq}=0.01\,M_{\odot}$ for
pulsars with a {He}~WD or a {CO}~WD companion, respectively.
In other words, to explain the observed period distribution of these two classes of recycled
pulsars we conclude that those pulsars with {CO}~WD companions typically have accreted less mass
by a factor of 6, compared to the pulsars with {He}~WD companions. This conclusion remains
valid for other scalings between the present spin period, $P$ and the initial spin period, $P_{\rm eq}$ --
even for assuming $P=P_{\rm eq}$.
We conclude from Fig.~\ref{fig:histo} that BMSPs with {CO}~WDs usually have evolved via early Case~B or Case~C RLO (the latter leading to CE evolution).
The reason is that these two paths involve mass transfer on a much shorter timescale, and hence 
lead to smaller amounts of transfered mass, leading to 
less effective recycling and thereby longer spin periods, as seen among the observed pulsars in Fig.~\ref{fig:histo} (see also Section~\ref{sec:CEspin}).
Furthermore, the BMSPs with {CO}~WDs also have relatively large values of $\dot{P}$.
This reflects larger values of $B$, which is expected since less mass was accreted in these systems, cf. Section~\ref{subsubsec:Bdecay}.
The only exception known from the description outlined above is \psr\ which is discussed in Section~\ref{sec:1614-2230}.\\  
For BMSPs with {He}~WD companions, we notice from the orbital period distribution in 
Fig.~\ref{fig:corbet3} (central panel) that the majority of these systems seem to have evolved through
Case~B RLO with an initial LMXB orbital period of $P_{\rm orb}> 1\,{\rm day}$ \citep[][and references therein]{ts99,prp02,tau11}. 

\subsection{Double neutron star systems}\label{subsec:NSNS}
In Fig.~\ref{fig:histo} we have also plotted the spin distribution of recycled pulsars with neutron star companions
and $P<100\,{\rm ms}$. These systems originate from high-mass {X}-ray binaries (HMXBs), e.g. \citet{tv06}. 
Their spin distribution is skewed to larger periods compared to the recycled pulsars with {CO}~WD companions.
This is expected since their massive stellar progenitors evolved even more rapidly than the donor stars of the IMXBs. 
It is generally believed that double neutron star systems evolved via a CE and spiral-in phase since the mass transfer in
HMXBs is always dynamically unstable -- see however \citet{bro95,dps06} for an alternative formation mechanism via a double-core scenario.

\subsection{Formation of recycled pulsars with $P_{\rm orb}>200\;{\rm days}$}\label{subsec:wLMXBs} 
As mentioned earlier (see also Fig.~\ref{fig:corbet3}), 
pulsars with $P_{\rm orb}>200\;{\rm days}$ all have slow rotational spins. This fact could be explained if they had accreted little mass.
If their progenitors were wide-orbit LMXBs they had experienced continuous super-Eddington mass transfer on a short timescale
of the order 10~Myr \citep[][]{ts99}. However, even if their accretion rates had been restricted to the Eddington limit
these accreting pulsars could still have received up to $0.3\,M_{\odot}$. 
This amount is 10 times larger than needed to spin~up a pulsar to 5~ms according to equation~(\ref{eq:deltaMfinalfit}). 
Therefore, to explain the slow spin periods of these wide-orbit recycled pulsars one may speculate that the
accretion process was highly inefficient due to, for example, enhanced accretion disk instabilities \citep{vpa96,dlhc99}. 
This could be related to the
advanced evolutionary stage of their donor stars on the red giant branch (RGB). These stars had deep convective envelopes 
which may lead to enhanced clump formation in the transfered material.\\
Alternatively, one may ask if these systems descended from wide-orbit IMXBs which underwent a CE and a mild spiral-in.
In this case their
companion stars should be fairly massive {CO}~WDs ($\ge 0.5\,M_{\odot}$, see fig.~1 in Paper~I).
However, quite a few of these wide-orbit recycled pulsars seem to 
have relatively low-mass {He}~WDs according to their mass functions. 
In fact it was pointed out by \citet{tau96,ts99,sfl+05} that some of those WD masses may even be significantly less massive than expected 
from the ($M_{\rm WD},\,P_{\rm orb}$)-relation which applies to {He}~WDs descending from LMXBs, 
although this analysis still has its basis in small numbers. 
It is therefore important to observationally determine the masses of the WDs in these wide-orbit pulsars systems and settle this evolutionary question. 

%%%%%%%%%%%%%%%%%%%%%%%%%%%%%%%%%%%%%%%%%%%%%%%%%%%%%%%%%%%%%%%%%%%%%%%%%%%%%%%%

\section{Spinning up post common envelope pulsars}\label{sec:CEspin} 
An interesting question is if our derived relation between pulsar spin and accreted mass (equation~\ref{eq:deltaMfinalfit})  
can also explain the observed spin periods of the BMSPs with {\it massive}
WD companions and short orbital periods, i.e. systems we expect to have evolved through a CE evolution. Let us consider the important
system PSR~J1802-2124 recently observed by \citet{fsk+10}. The nature of its $0.78\pm0.04\,M_{\odot}$ {CO}~WD companion, combined with an orbital period of 16.8~hr and a
post-accretion pulsar mass of only $1.24\pm0.11\,M_{\odot}$ makes this one of our best cases for a system which evolved via 
a CE and spiral-in phase. The present spin period of the pulsar is 12.6~ms. Assuming an initial equilibrium spin period of
10~ms we find from equation~(\ref{eq:deltaMfinalfit}) that $\Delta M_{\rm eq} \simeq 0.01\,M_{\odot}$ which is a much more acceptable value
for a system undergoing CE-evolution (compared to the $0.10\,M_{\odot}$ needed for spinning up a BMSP with
a He~WD companion to a spin period of a few ms).
However, one still has to account for the $0.01\,M_{\odot}$ accreted.
It is quite likely that this accretion occurred {\it after} the CE-phase and we now consider this possibility. 

\subsection{Case BB RLO in post common envelope binaries}\label{subsec:CaseBB} 
\begin{figure}
\begin{center}
  \includegraphics[width=0.34\textwidth, angle=-90]{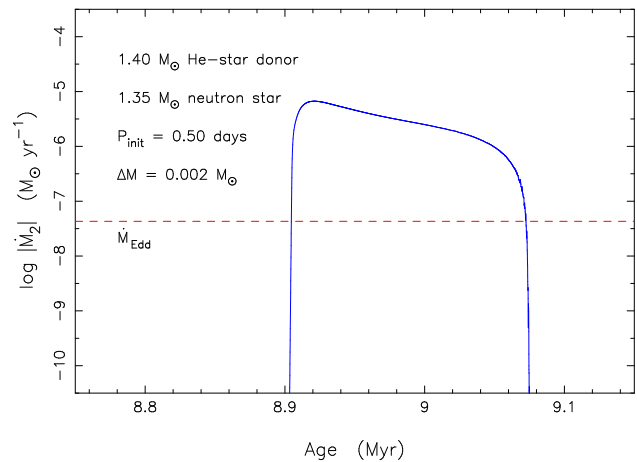}
  \caption[]{
    A model calculation of Case~BB RLO. The plot shows mass-transfer rate as a function of age for a $1.4\,M_{\odot}$ helium star donor.
    The initial orbital period is 0.50~days and the neutron star mass is $1.35\,M_{\odot}$.
    After the RLO the donor star settles as a $0.89\,M_{\odot}$ {CO}~WD orbiting the 
    recycled pulsar with an orbital period of 0.95~days.
    The mass-transfer phase only lasts for about 170.000~yr. Even if the neutron star only
    accretes $\sim\!0.002\,M_{\odot}$ (assuming an accretion efficiency of just 30\%)
    it is still able to spin up the pulsar to a spin period of 36~ms. If the neutron star
    accretes all matter at the Eddington limit it can be spun up to 14~ms. A comparison
    can be made with the observed spin periods of BMSPs with {CO}~WD companions in the bottom panel of Fig.~\ref{fig:corbet3}.
    }
\label{fig:HeNS}
\end{center}
\end{figure}

The total duration of the CE-phase itself is probably so short \citep[$<10^3\,{\rm yr}$, e.g.][]{pod01,pdf+11,ijc+12} 
that no substantial accretion
onto the neutron star is possible with respect to changing its rotational dynamics 
-- even if the envelope of the donor star is not ejected from the system until long after the in-spiral
of the secondary star which occurs on a dynamical timescale of a few $P_{\rm orb}$. 
Thus we are left with two possibilities: 1) the recycling of the pulsar occurs during 
the subsequent stable RLO of a naked
helium star following the CE (Case~BB RLO), or 2) the recycling occurs due to wind accretion from the naked helium
or carbon-oxygen core left behind by the donor star after it loses its envelope in the CE-phase.\\
In Fig.~\ref{fig:HeNS} we show one of our model calculations of Case~BB RLO. In our example we assumed a $1.40\,M_{\odot}$
helium star ($Y=0.98$) as the donor and a $1.35\,M_{\odot}$ neutron star as the accretor. The giant phase of a helium star is short lived
and in this case the mass transfer lasts for about 170.000~yr. (This time interval, however, is much longer than the CE-event.) 
Assuming the accretion onto the neutron star to be Eddington limited
will therefore lead to an accretion of only about $7\times 10^{-3}\,M_{\odot}$. Although this is a small amount, 
it is still sufficient to spin up the recycled pulsar to a period of 14~ms, which is even faster than the
observed median spin period of BMSPs with {CO}~WD companions (see Fig.~\ref{fig:histo}).
Assuming the accretion efficiency to be only 30~per~cent of $\dot{M}_{\rm Edd}$ would still spin up the pulsar to 36~ms. 
A helium donor star with a lower mass can even spin up the pulsar below 14~ms.
We found that for a $1.10\,M_{\odot}$ helium star donor the neutron star can accrete 50~per~cent more (resulting in 
$P_{\rm eq}=11\,{\rm ms}$) due to the increased nuclear timescale of the shell burning of a lower mass helium star. 
Hence, based on rotational dynamics we conclude that Case~BB RLO is a viable formation channel to explain most of the observed
BMSPs with {CO}~WD companions. A more systematic study of Case~BB RLO for various system configurations is needed in order to
find the minimum possible spin period of BMSPs with {CO}~WD companions formed via this formation channel. 

It should be noted that a pulsar may also be recycled via Case~BA RLO, i.e. from RLO which is initiated while
the companion star is still on the helium main sequence \citep[see][for detailed calculations]{dpsv02}. 
In this case the mass-transfer phase will last up to ten times longer, compared to Case~BB RLO, providing more than sufficient material
to recycle the pulsar effectively. However, in order to initiate RLO while the helium star donor is
still burning core helium on its main sequence requires a tight, fine-tuned interval of orbital separation following the common envelope.
Typical orbital periods needed are 1-2~hours. Since Case~BB RLO is initiated for a much wider interval of larger orbital periods
(up to several tens of days) we expect many more systems to evolve via Case~BB RLO compared to Case~BA RLO. 

\subsection{Wind accretion prior to Case~BB~RLO}\label{subsubsec:CaseBBwind} 
Finally, we tested if the neutron star could acquire any significant spin~up from wind accretion 
in the epoch between post-CE evolution and Case~BB RLO.
To produce {CO} or {ONeMg}~WD remnants via Case~BB RLO one would expect a typical helium star mass of $1.1-2.2\,M_{\odot}$
prior to the mass transfer. These stars have luminosities of $\log(L/L_{\odot})=2.5-3.3$ \citep{lan89} and typically spend less than 10~Myr 
on the {He}-ZAMS before they expand and initiate Case~BB RLO.  
Their mass-loss rates are of the order $\dot{M}_{\rm wind}\simeq 10^{-10\pm 0.6}\,M_{\odot}\,{\rm yr}^{-1}$ \citep{jh10}. 
Therefore, even if as much as 10~per~cent of this ejected wind mass was accreted onto the neutron star it would accrete
a total of at most $\sim 10^{-4}\,M_{\odot}$ prior to the RLO. 
We therefore conclude that wind accretion prior to Case~BB RLO is negligible 
for the recycling process of BMSPs with {CO}~WD companions.

\subsection{Will the progenitor binaries make it to Case~BB RLO?}\label{CaseBBmake}
Before drawing conclusions about the Case~BB RLO we should verify that the progenitor binary survives to follow this path.
In Paper~I we argued Case~BB RLO does not work for \psr\ because the envelope of the {CO}~WD progenitor star is too
tightly bound to be ejected during the spiral-in phase\footnote{We even considered a conservative Case~BB RLO evolution which is far 
from valid given that the mass-transfer rate is highly super-Eddington, see Fig.~\ref{fig:HeNS} above.}. 
Is that the case for all BMSPs? The answer is no. We find that
if the post-CE orbital period (i.e. the observed $P_{\rm orb}$ of a BMSP) 
is about 0.5~days or less then sufficient orbital energy will be released during spiral in
to energetically allow for a successful ejection of the envelope of the original $5-7\,M_{\odot}$ red giant donor star.
The total (absolute) binding energy of the envelope of such a star is $1.3-2.2\times 10^{48}\,{\rm ergs}$ which
is equivalent to an orbital energy corresponding roughly to $P_{\rm orb}\le 0.5\,{\rm days}$.
Since we must require $\Delta E_{\rm orb}>E_{\rm env}$ to avoid a merger 
we therefore conclude that Case~BB RLO is in principle a viable formation channel for BMSPs with {CO}~WD companions if 
these systems have short orbital periods. 
As argued earlier we consider PSR~J1802$-$2124 (which has $P_{\rm orb}=16.8\,{\rm hours}$) as a strong case for
a BMSP which evolved though a common envelope. Whether or not one can finetune the Case~BB RLO evolution to account for 
the recycling of this pulsar, or if the recycling was due to
wind accretion from the exposed post-CE {CO}~core of an AGB star or, thirdly, if accretion-induced collapse of
an {ONeMg}~WD was at work, is an interesting question to address. 

%%%%%%%%%%%%%%%%%%%%%%%%%%%%%%%%%%%%%%%%%%%%%%%%%%%%%%%%%%%%%%%%%%%%%%%%%%%%%%%%

\section{True ages of millisecond pulsars}\label{sec:trueages}
Knowledge of the true ages of recycled radio pulsars is important for comparing the observed population with the properties
expected from the spin-up theory outlined in Section~\ref{sec:spinup}. 
All radio pulsars lose rotational energy with age and the braking index, $n$ is given by \citep{mt77}:
\begin{equation} 
\label{eq:n}
    \dot{\Omega} \propto -\Omega ^n
\end{equation} 
which yields (for $n$ constant): $n\equiv \Omega\ddot{\Omega}/\dot{\Omega}^2$.
This deceleration law can also be expressed as: $\dot{P} \propto P^{2-n}$ and hence the
slope of a pulsar evolutionary track in the $P\dot{P}$--diagram is simply given by: $2-n$.
Depending on the physical conditions under which the pulsar spins down $n$ can take different values as mentioned earlier.
For example: 
\begin{equation}
  \begin{array}{llll}
    \mbox{gravitational wave radiation} & \mbox{\hspace{0.1cm} $n=5$} \\
    \mbox{B-decay or alignment or multipoles} & \mbox{\hspace{0.1cm} $n>3$} \\
    \mbox{perfect magnetic dipole} & \mbox{\hspace{0.1cm} $n=3$} \\
    \mbox{B-growth/distortion or counter-alignment} & \mbox{\hspace{0.1cm} $n<3$} \\
  \end{array}
\end{equation} 
The combined magnetic dipole and plasma current spin-down torque may also result in $n \ne 3$ \citep{cs06}. 
A simple integration of equation~(\ref{eq:n}) for a {\it constant} braking index ($n\ne1$)
yields the well-known expression:
\begin{equation} 
\label{eq:trueage}
   t=\frac{P}{(n-1)\dot{P}}\left[ 1-\left(\frac{P_0}{P}\right)^{n-1}\right]
\end{equation} 
where $t$ is the so-called true age of a pulsar, which had an initial spin period $P_0$ at time $t=0$.

\subsection{Isochrones in the $P\dot{P}$--diagram}\label{subsec:isochrones}
\begin{figure}
\begin{center}
  \includegraphics[width=0.35\textwidth, angle=-90]{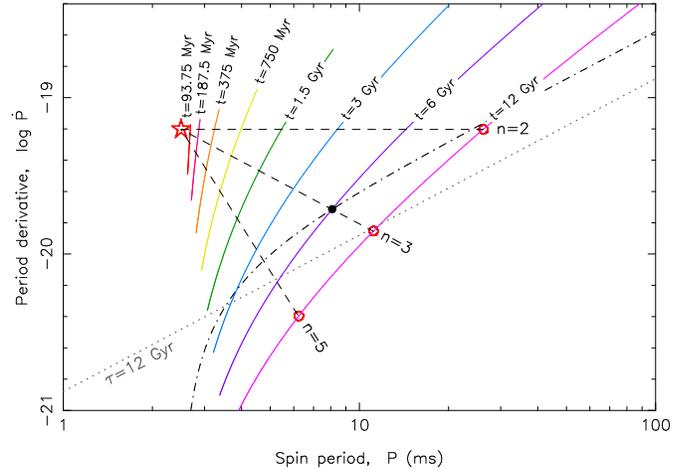}
  \caption[]{Isochrones of pulsar evolution for a pulsar with known initial $(P_0,\dot{P}_0)$ for different values 
             of a constant braking index, $n$. Along any given isochrone
             $n$ increases continuously from $n=1$ (top) to $n=15$ (bottom). The initial position of the recycled test pulsar is
             shown by a red star and the dashed lines show evolutionary tracks for $n=2$, 3 or 5, respectively.
             The dotted line yields a characteristic age, $\tau \equiv P/(2\dot{P})$ of $12\,{\rm Gyr}$. This line would intersect with the 
             12~Gyr isochrone exactly at $n=3$ if $P_0 \ll P$, which is not completely fulfilled. The dot-dashed isochrone line 
             is a general solution to equation~(\ref{eq:trueage}) for $n=3$, $t=6\,{\rm Gyr}$ and $P_0=2.5\,{\rm ms}$, but unknown $\dot{P}_0$.
    }
\label{fig:true_iso_single}
\end{center}
\end{figure}
\begin{figure*}
\begin{center}
  \includegraphics[width=0.40\textwidth, angle=-90]{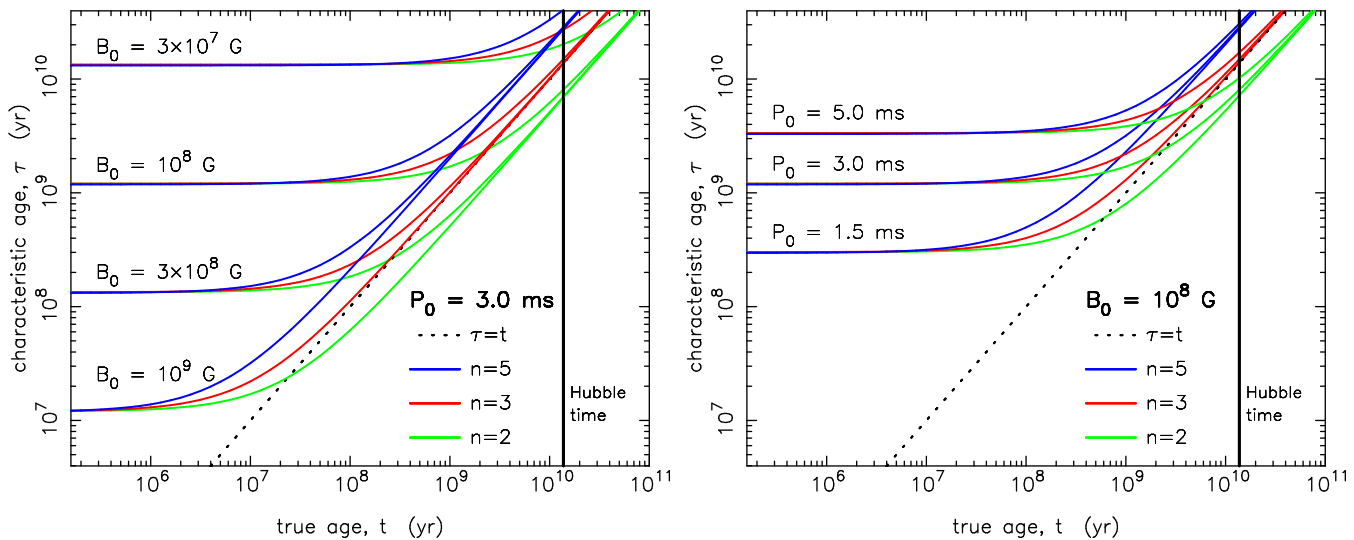}
  \caption[]{Evolutionary tracks of characteristic ages, $\tau$ calculated as a function of true ages, $t$ for recycled pulsars with 
             a constant braking index of $n=2$, $n=3$, and $n=5$. In all cases for $n$ we assumed a constant $M=1.4\,M_{\odot}$, $\sin \alpha =1$, a constant
             moment of inertia, $I$ and 
             for $n=5$ we also assumed a constant ellipticity, $\varepsilon \neq 0$.  
             In the left panel we assumed in all cases an initial spin period of $P_0=3.0\,{\rm ms}$ and varied the value of the initial
             surface magnetic flux density, $B_0$. In the right panel we assumed in all cases $B_0=10^8\,{\rm G}$
             and varied $P_0$. The dotted line shows a graph for $\tau=t$ and thus
             only pulsars located on (near) this line have characteristic ages as reliable age indicators. We notice that recycled pulsars
             with small values of $\dot{P}_0$, resulting from either small values of $B_0$ (left panel) 
             and/or large values of $P_0$ (right panel), tend to have $\tau \gg t$, even at times exceeding 
             the age of the Universe. In all degeneracy splittings of the curves the upper curve (blue) corresponds to $n=5$, the central curve (red) corresponds
             to $n=3$ and the lower curve (green) corresponds to $n=2$. The asymptotic behavior of the curves can easily be understood
             from equation~(\ref{eq:tau_asymp}) -- see text for explanations.           
    }
\label{fig:ages}
\end{center}
\end{figure*}
Unfortunately, one cannot directly use the equation above to obtain evolutionary tracks
in the $P\dot{P}$--diagram (even under the assumption of a constant $n$
and a known initial spin period, $P_0$). 
For chosen values of $t$, $n$ and $P_0$ one can find a whole family of solutions $(P,\dot{P}$) to be plotted
as an isochrone, see the dot-dashed line in Fig.~\ref{fig:true_iso_single} \citep[and see also][]{kt10}. However, there is only one point
which is a valid solution for a given pulsar and the above equation does reveal which point is correct.
The problem is that the variables in equation~(\ref{eq:trueage}) are not independent and we do not know a~priori 
the initial spin period derivative, $\dot{P}_0$.
The evolution of the spin period is a function of both $t$, $n$, the initial period and its time derivative, i.e. $P(t,n,P_0,\dot{P}_0)$. 
Here, for simplicity, we assume $n$ to be constant.
To determine $\dot{P}_0$ one must make assumptions about the initial 
surface magnetic flux density, $B_0$, the initial magnetic inclination angle, $\alpha _0$ and the initial ellipticity, $\varepsilon _0$ 
(the geometric distortion which is relevant for rotational deceleration due to gravitational wave radiation). Once $\dot{P}_0$ is known,
one can combine equation~(\ref{eq:trueage}) with the deceleration law, e.g. $\dot{P}P^{n-2}=const.$ (equation~\ref{eq:n}), 
and solve by integration for the evolution
of a given pulsar. This way one can produce isochrones, for example as a function of the braking index for pulsars with
the same given value of $\dot{P}_0$.\\
In Fig.~\ref{fig:true_iso_single} we have plotted pulsar isochrones for a pulsar with $P_0=2.5\,{\rm ms}$ and $\dot{P}_0=6.26\times 10^{-20}$
(e.g., corresponding to $B_0=1.15\times 10^8\,{\rm G}$ for $M=1.4\,M_{\odot}$, $\sin \alpha _0=1$ and $\varepsilon _0=0$) by varying the value of the
braking index such that $1 \le n \le 15$. Our numerical calculations were
confirmed by analytic calculations for the special cases where $n=2$,~3~or~5 (see small circles at the $t=12\,{\rm Gyr}$ isochrone).
The black dot indicates where the two 6~Gyr isochrones, calculated for $n=3$ or $\dot{P}_0=6.26\times 10^{-20}$, respectively,
cross each other for a common solution and also in conjunction with the intersection of the evolutionary track calculated
for the same values of $n$ and $\dot{P}_0$.

\subsection{Characteristic versus true ages of MSPs}\label{subsec:tau_t}
Introducing the characteristic age of a pulsar, $\tau \equiv P/(2\dot{P})$ one finds
the relation between $\tau$ (the observable) and $t$:
\begin{equation}
 \displaystyle \log\tau= \log t + \log\left(\frac{n-1}{2}\right) -\log \left(1-(\frac{P_0}{P})^{n-1}\right)
\end{equation} 
for evolution with a constant braking index, $n$.
The asymptotic version of this relation (as $t\rightarrow\infty$ and $P\gg P_0$) is given by:
\begin{equation}\label{eq:tau_asymp}
 \log\tau=\left\{ \begin{array}{lll}
   \displaystyle \log t + \log 2 & \mbox{\hspace{0.1cm}for $n=5$} \\
   \displaystyle \log t          & \mbox{\hspace{0.1cm}for $n=3$} \\
   \displaystyle \log t - \log 2 & \mbox{\hspace{0.1cm}for $n=2$} \\
           \end{array}
         \right.  
\end{equation} 
In Fig.~\ref{fig:ages} we have plotted $\tau$ as a function of $t$ on log-scales for braking indices of $n=2$, 3 and 5, respectively. 
Initially, when $t$ is small and $P\simeq P_0$,
$\tau$ is always greater than $t$. After a certain timescale $\tau$ can either remain larger or become smaller than $t$, depending on $n$.
For $n=5$ ($n>3$) we always have $\tau > t$, for $n=3$ we have $\tau \ge t$ and only for $n=2$ ($n<3$) we have the possibility that
$\tau$ can be either greater or smaller than $t$. Having $n=2$ corresponds to $\dot{P}$ being a constant, and
solving for $\tau =t$ yields $P=2P_0$ and $t=P_0/\dot{P}$.
We notice that for a reasonable interval of braking index values $2\le n \le 5$ the observed characteristic age will never deviate
from the true age by more than a factor of two when $P \gg P_0$.
However, to reach $P\gg P_0$ may take several Hubble timescales if $B_0$ is low.

The characteristic age at birth (after recycling) is given by: $\tau _0 = P_0/(2\,\dot{P}_0)$ and combining with equation~(\ref{eq:Bspitkovsky}) we get:
\begin{equation} 
\label{eq:tauzero}
   \tau _0=\frac{P_0^2\,k^2}{2B_0^2}
\end{equation} 
where the constant $k=9.2\times 10^{18}\,{\rm G\,s}^{-1/2}$ for $\sin \alpha _0 =1$ and $M=1.4\,M_{\odot}$.
This expression verifies the well-known result that MSPs that are born relatively slowly spinning (large $P_0$ value) 
and/or have a relatively weak initial $B$-field (small $B_0$ value)
are also those MSPs which are born with characteristic ages which differ the most from their true ages as shown
in Fig.~\ref{fig:ages} \citep[see also][]{kt10}.
In other words, those pulsars with smallest values of $\dot{P}_0$ for a given $B_0$ or $P_0$ are those
with $\tau _0 \gg t$ (which is hardly a surprise given that $\tau \equiv P/(2\dot{P})$).
The extent to which they continue evolving with $\tau \gg t$ depends on their initial conditions ($P_0, B_0$) and well as $n$.
Notice, for recycled pulsars $n$ is not measurable and for this reason its value remains unknown.  

\subsection{Comparison with observations}\label{subsec:isochrones_ppdot}
\subsubsection{Kinematic corrections to $\dot{P}$}\label{subsubsec:Pdotcorr}
In a discussion of the ages and the evolution of millisecond pulsars in the
$P\dot{P}$--diagram kinematic corrections must be included when considering the observed values of $\dot{P}$ ($\dot{P}_{\rm obs}$). 
The kinematic corrections to the intrinsic $\dot{P}$ ($\dot{P}_{\rm int}$) are caused by acceleration due to proper motion of
nearby pulsars \citep{shk70} and to vertical ($a_Z$) and 
differential rotational acceleration in our Galaxy ($a_{\rm GDR}$).
The total corrections are given by:
\begin{equation} 
\label{eq:Pdotcorr}
  \left( \frac{\dot{P}_{\rm obs}}{P} \right) =  \left( \frac{\dot{P}_{\rm int}}{P} \right)  +
             \left( \frac{\dot{P}_{\rm shk}}{P} \right)  + \frac{a_Z}{c} + \frac{a_{\rm GDR}}{c}
\end{equation} 
and following \citet{dt91} and \citet{wdkk+00} we can express these corrections as: 
\begin{eqnarray} 
\label{eq:Pdotcorr2}
  \left( \frac{\dot{P}_{\rm obs}}{P} \right) &=& \left( \frac{\dot{P}_{\rm int}}{P} \right)  +
             \frac{\mu^2\,d}{c}  - \frac{a_Z\,\sin b}{c} \\ \nonumber
          &-& \frac{v_0^2}{cR_0}\left[\cos l + \frac{(d/R_0) - \cos l}{1+(d/R_0)^2-2(d/R_0)\cos l} \right]
\end{eqnarray} 
where $d$ is the distance to the pulsar, $\mu$ is the proper motion (related to the transverse velocity, $v_\bot=\mu d$),
$a_Z$ is the vertical component of Galactic acceleration (e.g., from the Galactic potential model
of \citet{kg89}), $l$ and $b$ are the Galactic coordinates of the pulsar, $R_0=8.0\,{\rm kpc}$ and $v_0=220\,{\rm km\,s^{-1}}$ represent the
distance to the Galactic centre and the orbital velocity of the Sun, and $c$ is the speed of light in vacuum.
The corrections to $\dot{P}$ due to Galactic vertical and differential rotational accelerations are typically quite small (a few $10^{-22} \ll \dot{P}_{\rm int}$)
and can be ignored, except in a few cases (see below).

\subsubsection{Evolutionary tracks and true age isochrones}\label{subsubsec:PPdot_isochrones}
\begin{figure*}
\begin{center}
  \includegraphics[width=0.60\textwidth, angle=-90]{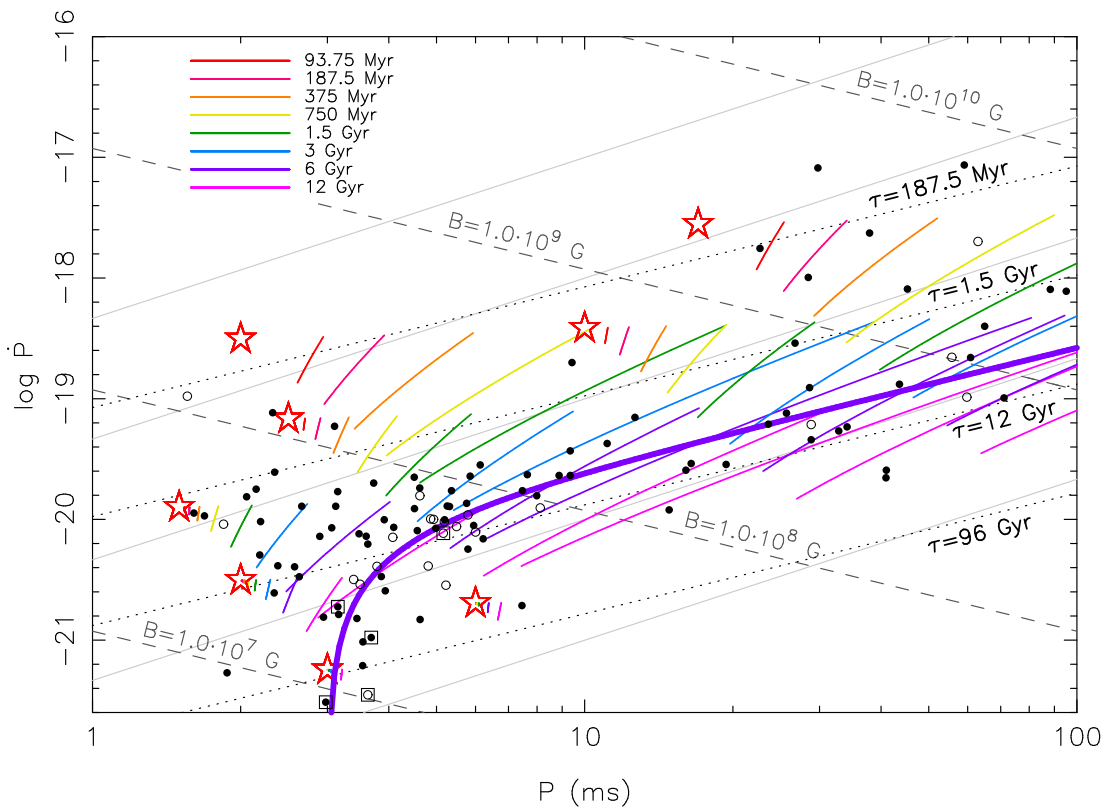}
  \caption[]{Isochrones of eight hypothetical recycled pulsars born at the locations of the red stars. 
             The isochrones were calculated for different values of the braking index, $2\le n\le 5$
             (see Fig.~\ref{fig:true_iso_single} for details). Also plotted are inferred $B$-field values (dashed lines) and
             characteristic ages, $\tau$ (dotted lines). 
             The thin grey lines are spin-up lines with $\dot{M}/\dot{M}_{\rm Edd}=1,\,10^{-1},\,10^{-2},\,10^{-3}$ and $10^{-4}$ (top to bottom,
             and assuming $\sin \alpha =\phi =\omega_{\rm c}=1$).
             In all calculations we assumed a pulsar mass of $1.4\,M_{\odot}$.
             It is noted that the majority of the observed population is found near the isochrones for $t=3-12\,{\rm Gyr}$, as expected.
             The fat, solid purple line indicates an example of a $t=6\,{\rm Gyr}$ isochrone
             for pulsars with any value of $\dot{P}_0$, but assuming $P_0=3.0\,{\rm ms}$ and $n=3$. It is seen how the banana shape of
             such a type of an isochrone fits very well with the overall distribution of observed pulsars in the Galactic disk.
             Binary pulsars are marked with solid circles and isolated pulsars are marked with open circles. 
             All values of $\dot{P}$ have been corrected for the Shklovskii effect using data from the {\it ATNF Pulsar Catalogue}.
             Pulsars in squares have been corrected for both a reduced distance and acceleration effects in the Galaxy.
             Pulsars with very small values of $\dot{P}$ are basically born on location in this diagram since their evolutionary
             timescale exceeds the Hubble~time. Pulsars with relative small values of $P$ and large values of $\dot{P}$ have the
             potential to constrain white dwarf cooling models -- see text for discussion.
    }
\label{fig:true_iso_all}
\end{center}
\end{figure*}

In order to investigate if we can understand the distribution of MSPs in the $P\dot{P}$--diagram we have traced the evolution
of eight hypothetical, recycled MSPs with different birth locations. In each case we traced the evolution for $ 2\le n \le 5$ and plotted
isochrones similar to those introduced in Fig.~\ref{fig:true_iso_single}. The results are shown in Fig.~\ref{fig:true_iso_all}
together with observed data. 
All the measured $\dot{P}$ values have been corrected for kinematic effects, as described above. 
If the transverse velocity of a given pulsar is unknown we used a value of $67\,{\rm km\,s}^{-1}$ which we found
to be the median value of the 49 measured velocities of binary pulsars\footnote{If we add to this sample the measured velocities
of 13 isolated pulsars which show strong signatures of being recycled ($P<100\,{\rm ms}$ and $\dot{P}<10^{-16}$) the median velocity becomes
$69\,{\rm km\,s}^{-1}$, which would make the Shklovskii corrections slightly larger and thus $\dot{P}_{\rm int}$ somewhat smaller.}.
In five cases (PSRs: J1231$-$1411, J1614$-$2230, J2229+2643, J1024$-$0719, J1801$-$1417, marked in squares in Fig.~\ref{fig:true_iso_all}) 
we obtain $\dot{P}_{\rm int}<0$ which is not physically possible. The reason is probably an overestimate of the pulsar distance. In those cases
we have recalculated $\dot{P}_{\rm int}$ assuming only half the distance and included the corrections due to  
Galactic vertical and differential rotational acceleration.\\
Three main conclusions can be drawn from this diagram:\\ 
1) The overall distribution of observed pulsars
follows nicely the banana-like shape of an isochrone with multiple choices for $\dot{P}_0$ (or $B_0$), see fat purple line. 
The chosen values of $P_0=3.0\,{\rm ms}$,
$n=3$ and $t=6\,{\rm Gyr}$ are just for illustrative purposes only and not an attempt for a best fit to the observations.
Fitting to one curve would not be a good idea given that MSPs are born with different initial spin periods, which depend on their accretion history,
and given that the pulsars have different ages. The spread in the observed population is hinting that recycled pulsars are born at many different locations
in the $P\dot{P}$--diagram.\\
2) The far majority of the recycled pulsars seem to have true ages
between 3 and 12~Gyr, as expected since the population accumulates and the pulsars keep emitting radio waves for a Hubble time.\\
3) Pulsars with small values of the period derivative $\dot{P}\simeq 10^{-21}$ hardly evolve at all in the diagram over a Hubble time. This is 
a trivial fact, but nevertheless important since it tells us that these pulsars were basically born with their currently observed
values of $P$ and $\dot{P}$ \citep[first pointed out by][]{ctk94}. 
In this respect, it is interesting to notice PSR~J1801$-$3210 \citep[recently discovered by ][]{bates+11} which
must have been recycled with a relatively slow birth period, $P_0 \sim \! 7\;{\rm ms}$ despite its low B-field $<10^8\;{\rm G}$
-- see Fig.~\ref{fig:spinupline} for its location in the $P\dot{P}$--diagram. 

One implication of the third conclusion listed above is that some radio MSPs must have been born with 
very small values of $B_0\simeq 1\times 10^7\,{\rm G}$,
if the inferred $B$-fields using equation~(\ref{eq:Bspitkovsky}) are correct. 
Given their weak B-fields such sources would most likely not be able to channel the accreted matter sufficiently to become
observable as AXMSPs \citep{wv98}.
One case is the 1.88~ms radio pulsar
J0034$-$0534 \citep{bhl+94} which has an observed $\dot{P}=4.97\times 10^{-21}$. However, this pulsar has a transverse velocity
of $146\,{\rm km\,s}^{-1}$ and correcting for the Shklovskii effect yields an intrinsic period derivative of
only $\dot{P}_{\rm int}=5.36\times 10^{-22}$. These values result in $B=9\times 10^6\,{\rm G}$ 
(for $M=1.4\,M_{\odot}$ and $\alpha = 90^{\circ}$) and a characteristic age, $\tau = 55\,{\rm Gyr}$.
Once again we see that $\tau$ is not a good measure of the true age of a pulsar.
If this pulsar has any gravitational wave emission
($\varepsilon \ne 0$) its derived surface $B$-field would be even smaller.
We also notice a few pulsars which seem to have been recycled with $\dot{M}<10^{-3}\,\dot{M}_{\rm Edd}$
(approaching $10^{-4}\,\dot{M}_{\rm Edd}$). Their progenitors
could be the so-called weak~LMXBs which are low-luminosity LMXBs \citep{vdk06}. It is also quite possible that their
position in the $P\dot{P}$--diagram was strongly shifted during the RLDP \citep[see fig.~2 in][]{tau12}
so that their average mass-accretion rate could have been larger prior to the RLDP.
Alternatively, these MSPs demonstrate that $\phi >1$ and $\omega _{\rm c}<1$ which could move 
the spin-up line substantially downwards, as demonstrated in Fig.~\ref{fig:spinupline}.  

It is obvious that a different sample of recycled pulsar birth locations in Fig.~\ref{fig:true_iso_all} 
would lead to somewhat different isochrones. However, we believe the qualitative interpretation
is trustworthy and encourage further detailed statistical population analysis to be carried out \citep[see][]{kt10}.

It is worth pointing out the difference in evolutionary timescales for the progenitor systems of BMSPs with {He}~WDs or {CO}~WDs, respectively.
The IMXBs, leading mainly to BMSPs with {CO}~WD companions, 
evolve on nuclear timescales of typically 100$-$300~Myr. Hence, BMSPs with {CO}~WDs could have true
ages from zero to up to about 12~Gyr (assuming the first progenitor stars in our Galaxy formed $\sim\!1\,{\rm Gyr}$ after Big~Bang). The LMXBs, on the other hand,
could easily have ages of 10~Gyr {\it before} they form, which is the typical timescale for a $1\,M_{\odot}$ star to evolve into 
a sub-giant which fills its Roche-lobe in an orbit with $P_{\rm orb}$
larger than a few days. One may then think that BMSPs with {He}~WD companions must be much younger than BMSPs with {CO}~WD companions (the BMSP age
being calculated from after the RLDP). However, some LMXBs may have donor stars up to $2\,M_{\odot}$ and these stars evolve on a timescale
of only $\sim\!1.5\,{\rm Gyr}$. Hence, BMSPs with {He}~WD companions can have almost similar true (post recycling) ages as BMSPs
with {CO}~WD companions.

\subsubsection{PSR~B1937+21 and PSR~J0218+4232: two recently fully recycled pulsars?}\label{subsubsec:twoPSRs}
Although recycled pulsars with high values of $\dot{P}$ evolve fast across the $P\dot{P}$--diagram, we should expect
to observe a few fully recycled pulsars with $\dot{P}\simeq 10^{-19}$.
The only two cases in the Galactic disk known so far are PSR~B1937+21 \citep{bkh+82} 
and PSR~0218+4232 \citep{ndf+95}, see Fig.~\ref{fig:spinupline}. 
The latter is a 2.32~ms pulsar ($\dot{P}_{\rm int}=7.66\times 10^{-20}$) which not only has a small
characteristic age ($\tau = 480\,{\rm Myr}$) but also must have a fairly young {\it true} age. Even if we assume it evolved with a constant
value of $\dot{P}$ (and disregard the possibility of $n>2$ which would have resulted in a larger initial value of $\dot{P}_0$ and hence
a younger age) we find $t=132-546\,{\rm Myr}$ when assuming $P_0$ is in the interval 1--2~ms. 
As expected, this system hosts a white dwarf companion star which is still hot enough to be detected.
Interestingly enough, all the cooling models applied by \citet{bvk03} to this white dwarf seem to indicate a somewhat larger age (see their fig.~4). This discrepancy
has the intriguing consequence that either their applied cooling models are overestimating the age of the white dwarf, or this pulsar
evolved with $n<2$ and evolved {\it upward} in the $P\dot{P}$--diagram.\\

PSR~B1937+21 is the first MSP discovered and is not only young, like PSR~J0218+4232, but also isolated.
This fast spinning pulsar (1.56~ms) has $\dot{P}_{\rm int}=1.05\times 10^{-19}$ which yields a characteristic age of
$\tau = 235\,{\rm Myr}$. We find upper limits ($n=2$) for the true age to be in the interval $t=78-229\,{\rm Myr}$
when assuming $P_0$ is in the interval 0.8--1.3~ms. MSPs (or any pulsar with a low $\dot{P}$) are not believed to be formed directly from 
supernova explosions. They must interact with a companion star to spin~up and decrease their B-field by a large amount. Thus
it is interesting that this pulsar was able to evaporate its
companion star within a timescale of (a few) $10^8\,{\rm yr}$ 
-- unless it evolved from an ultra-compact {X}-ray binary possibly initially leaving a planet around it \citep{vnvj12b}. 

\subsubsection{PSR~J1841+0130: a young mildly recycled pulsar}\label{subsubsec:1841}
PSR~J1841+0130 is the 29~ms binary pulsar discussed earlier in Section~\ref{subsubsec:1841wd}. It has an unusually high value of $\dot{P}$, close to $10^{-17}$,
which reveals that it is young and places it close to the spin-up line for $\dot{M}=\dot{M}_{\rm Edd}$ (assuming $\sin \alpha = \phi = \omega _c =1$), 
see Fig.~\ref{fig:spinupline}. The characteristic age of PSR~J1841+0130 is 58~Myr. An upper limit to its
true age is 77~Myr, which is found by assuming $n=2$ and $P_0=10\,{\rm ms}$ ($P_0$ cannot have been much 
smaller than 10~ms under the above mentioned assumptions regarding the spin-up line). If we assume $n=3$, the upper limit for the
true age is 51~Myr. The WD companion is therefore hot enough to yield spectroscopic information that will reveal its true nature --
i.e. whether this is a {CO}~WD or a {He}~WD, as discussed in Section~\ref{subsubsec:1841wd}. 
According to the dispersion measure, $DM=125\,{\rm cm}^{-3}\,{\rm pc}$, of this pulsar its distance is about 3~kpc which might
make such observations difficult.
 
%%%%%%%%%%%%%%%%%%%%%%%%%%%%%%%%%%%%%%%%%%%%%%%%%%%%%%%%%%%%%%%%%%%%%%%%%%%%%%%%

\section{Spinning up PSR~J1614$-$2230}\label{sec:1614-2230}
Recent Shapiro delay measurements of the radio millisecond pulsar J1614$-$2230 \citep{dpr+10} allowed a precise mass determination
of this record high-mass neutron star and its white dwarf companion.
A few key characteristic parameters of the system are shown in Table~\ref{table:param}.
%-------------------------------------------------------------------------------
\begin{table}
\center
\caption{Selected physical parameters of the binary millisecond pulsar J1614$-$2230 
         \citep[data taken from ][]{dpr+10}.}
\begin{tabular}{lr}
%\hline {Parameter} & {value} \\ \noalign{\mallskip}
\hline {Parameter} & {value} \\ 
\hline 
\noalign{\smallskip} 
            Pulsar mass & $1.97\pm 0.04\,M_{\odot}$ \\
            White dwarf mass & $0.500\pm 0.006\,M_{\odot}$ \\
            Orbital period & $8.6866194196(2)\;\rm{days}$ \\
%           Projected pulsar semimajor axis & $11.2911975\;\rm{light~sec}$ \\
            Orbital eccentricity & $1.30\pm 0.04 \times 10^{-6}$ \\
%           Inclination angle & $89.17\pm 0.02\;\rm{deg.}$ \\
%           Dispersion-derived distance & $1.2\;\rm{kpc}$ \\
            Pulsar spin period & $3.1508076534271\;\rm{ms}$ \\
            Period derivative & $9.6216\times 10^{-21}$ \\
\noalign{\smallskip} 
\hline
\end{tabular}
\label{table:param}
\end{table}
%-------------------------------------------------------------------------------
In Paper~I we discussed the binary evolution which led to the formation of this system \citep[see also][]{lrp+11}.  
For estimating the amount of necessary mass accreted in order to recycle \psr\ we can
apply equation~(\ref{eq:deltaMfinalfit}) and assuming, for example, an initial spin period of 2.0~ms following the RLO.  
We find that $\Delta M_{\rm eq} = 0.11\,M_{\odot}$. This result is in fine accordance 
with our Case~A calculation in Paper~I where a total of 
$0.31\,M_{\odot}$ is accreted by the neutron star. (Recall that $\Delta M_{\rm eq}$ is a minimum value for ideal, efficient spin~up.) 
For the Case~C scenario it is difficult to reconcile the required accretion of $0.11\,M_{\odot}$ with the very small amount
which is expected to be accreted by the neutron star evolving through a CE-phase (usually assumed to be $\ll 10^{-2}\,M_{\odot}$). 
Furthermore, in Paper~I we argued that Case~BB RLO (stable RLO from a naked helium star following a CE) was not an option for this system.\\
The next check is to see if the spin-relaxation time scale was shorter than the mass-transfer timescale for \psr. 
In order to estimate $t_{\rm torque}$ we must have an idea of the B-field strength during the RLO. Since we have shown in Paper~I that
the mass-transfer rate in the final phase AB was near (slightly above) the Eddington limit, the value of $B$ prior to phase~AB must have been somewhat larger
than its current estimated value of $\sim 8.4\times 10^7\,{\rm G}$, according to equation~(\ref{eq:Bspitkovsky}).  
Using $B=4-10\times 10^8\,{\rm G}$ we find $t_{\rm torque}\simeq 2-9\,{\rm Myr}$. Since this timescale is shorter than the 
duration of RLO ($\sim\!10\,{\rm Myr}$ for phase~AB, see Paper~I) we find that indeed it was possible for \psr\ to spin up to its equilibrium period.\\
Based on both binary evolution considerations (Paper~I) and
the spin dynamics (this paper) we conclude that the Case~A scenario is required to explain the existence of \psr.
From Fig.~\ref{fig:histo} we notice that \psr\ is an anomaly among the population of BMSPs with {CO}~WD companions -- it has an unusual rapid spin.
This is explained by its formation via a stable and relatively long lasting IMXB Case~A RLO (Paper~I) which is, apparently, 
not a normal formation scenario for these systems.  

\begin{figure}
\begin{center}
  \includegraphics[width=0.35\textwidth, angle=-90]{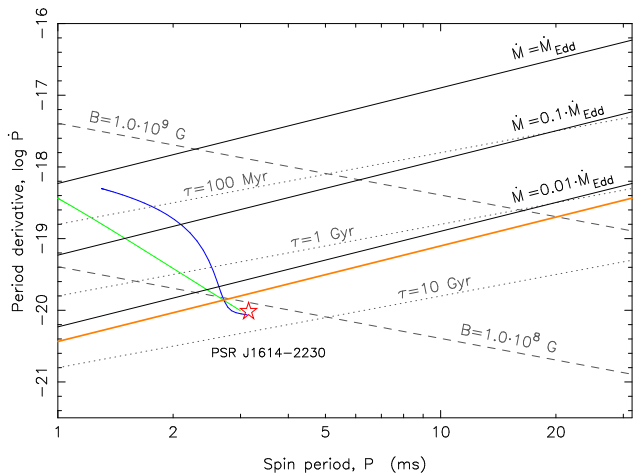}
  \caption[]{
             Simple models of how \psr\ could have evolved from a birth near a spin-up line with $\dot{M}=\dot{M}_{\rm Edd}$ 
             to its currently observed location, shown by the red star, within 2~Gyr. 
             The orange line indicates the location of the
             $\dot{M}=\dot{M}_{\rm Edd}$ spin-up line if $\alpha =10^{\circ}$, $\phi =1.4$ and $\omega _c=0.25$. 
             \psr\ is located close to this line which suggests it was born (recycled) close to it. 
             The three black solid lines were calculated by using
             $\alpha=90^{\circ}$, $\phi=1$ and $\omega _c=1$. 
             If \psr\ was born near such a spin-up line with $\dot{M}=\dot{M}_{\rm Edd}$ the need for subsequent spin-down after its birth
             would be severe and we have evolved a couple of toy models for this alternative: 
             The green model represents gravitational wave radiation
             and the blue model represents a phase of enhanced torque decay (see text). 
    }
\label{fig:ppdot_evol}
\end{center}
\end{figure}

\subsection{Evolution of \psr\ in the $P\dot{P}$--diagram}\label{subsec:ppdot-evol} 
We have demonstrated in Paper~I that \psr\ mainly accreted its mass during the final phase (AB) of mass transfer.
As mentioned above, 
in this phase the mass-transfer rate was high enough that the accretion onto the neutron star was limited by the Eddington limit,
i.e. $\dot{M}=\dot{M}_{\rm Edd}$.
A natural question to address (see also \citet{bk11}) is then how \psr\ evolved to the present location in the $P\dot{P}$--diagram which, 
at first sight, is even below the spin-up line expected for $\dot{M}=10^{-2}\,\dot{M}_{\rm Edd}$. 
However, one must bear in mind the dependency of the parameters $\alpha$, $\phi$ and $\omega _c$ when discussing
the location of the spin-up line (cf. Section~\ref{subsec:spinupline}).

If \psr\ has a magnetic inclination angle, $\alpha < 90^{\circ}$, and in case $\phi\approx 1.4$ and $\omega_c\approx 0.25$, 
then the problem is solved since the
$\dot{M}=\dot{M}_{\rm Edd}$ spin-up line (see orange line in Fig.~\ref{fig:ppdot_evol}) moves close to the current position of \psr.
Therefore we do not see any reason to question the recycling model of BMSPs,
based on the observations of the \psr\ system, as proposed by \citet{bk11}. 
However, if $\phi \approx \omega_c \approx 1$ and $\alpha \approx 90^{\circ}$ then we have to consider an alternative explanation
which we shall now investigate.  

It is well-known that the rotational evolution of millisecond pulsars is rather poorly understood
since the braking index only has been measured accurately for a few young, non-recycled pulsars.
As briefly mentioned in Section~\ref{sec:trueages} 
there is a variety of mechanisms which can influence the braking index of a pulsar. For example, the geometry of the B-field, the decay of the B-field, 
a changing magnetic inclination angle (i.e. between the spin axis and the B-field axis of the neutron star), the dynamics of the current flow 
in the pulsar magnetosphere and gravitational wave emission, e.g.
\citet{mt77} and \citet{cs06}.
We have made a simple toy model for two of these mechanisms to show that \psr\ could in principle have accreted with $\dot{M}=\dot{M}_{\rm Edd}$
for $\sin \alpha \approx \phi \approx \omega_c \approx 1$ 
and subsequently evolve to its presently observed values of $P$ and $\dot{P}$ 
(near the corresponding $\dot{M}=10^{-2}\,\dot{M}_{\rm Edd}$ spin-up line)
after the radio pulsar turned on. 
In the first model we assumed pulsar spin-down as a result
of gravitational wave radiation, which corresponds to $n=5$. 
The emission of gravitational waves is caused by a time-varying quadrupole moment which arises, for example, 
if the accretion onto the neutron star created non-axisymmetric moments of inertia relative to the spin axis \citep{st83}.  
In the second model we assumed an enhanced torque decay ($n>3$) for a limited amount of time following the recycling phase.
Such enhanced torque decay could be caused by either alignment between the B-field axis and the spin axis of the pulsar  
(but not complete alignment, in which case no pulsar would be seen) or further decay of the surface B-field through some unspecified mechanism.
In our calculation for the second model we assumed that effectively $B$ or $\sin\alpha$ decreased by a factor of about 
five on a decay timescale of 200~Myr.
In Fig.~\ref{fig:ppdot_evol} we show that the present position in $P\dot{P}$--diagram can be reached from such toy models
within 2~Gyr, which is in accordance
with the cooling age of the {CO}~WD inferred by \citet{bk11}. 
Notice, if \psr\ accreted $0.31\,M_{\odot}$, as suggested in Paper~I, it would in principle be possible for it to spin~up to a period of
0.92~ms (disregarding possible braking torque effects due to gravitational wave radiation, \citet{bil98}) 
and thus the initial spin periods in our toy models are justified. 
However, we emphasize again that it is much less
cumbersome to explain the current location of \psr\ in the $P\dot{P}$--diagram simply by finetuning, in particular, $\phi$ and $\omega_c$.
Therefore, future observations of systems like \psr\ may help to constrain the disk-magnetosphere interactions.

%%%%%%%%%%%%%%%%%%%%%%%%%%%%%%%%%%%%%%%%%%%%%%%%%%%%%%%%%%%%%%%%%%%%%%%%%%%%%%%%

\section{Neutron star birth masses}\label{sec:NSmass}
In Paper~I, we concluded that the neutron star in \psr\
must have been born in a supernova explosion (SN) with a mass of $1.7\pm0.15\,M_{\odot}$ if it subsequently evolved through an
{X}-ray phase with Case~A RLO. Based on evolutionary considerations we also found
an alternative scenario where the post-SN system could have evolved via Case~C RLO leading to a CE. However, we also argued
that the CE scenario was less probable given the difficulties in the required evolution from 
the ZAMS to the {X}-ray phase. In this paper we rule out the CE scenario completely in view of
our analysis of the rotational dynamics. The in-spiral of the neutron star in a CE occurs 
on such a short timescale that no significant spin~up is possible, and wind accretion from the proto~WD
cannot recycle the pulsar to a spin period of 2$-$3~ms which requires accretion of $0.1\,M_{\odot}$. Recycling
via a Case~BB RLO, following CE evolution initiated while the progenitor of the donor star was on the RGB,
is also not possible for \psr\ since its orbital period is so large that its post-CE orbit must have been wide too and thus there could not
have been released enough orbital energy during the in-spiral to eject the envelope.\\
Does such a high birth mass of $1.7\,M_{\odot}$ significantly exceed the neutron star birth masses in previously discovered radio pulsar systems?
Below we test if other pulsars may have been born in a SN with a similarly high mass.

The interval of known radio pulsar masses ranges from $1.17\,M_{\odot}$ in
the double neutron star binary PSR~J1518+4909 \citep[3-$\sigma$ upper limit,][]{jsk+08}
to $1.97\,M_{\odot}$ in \psr, discussed in this paper. Prior to the measurement of \psr\ the highest mass
inferred for the birth mass of a radio pulsar was only
$1.44\,M_{\odot}$ (PSR~1913+16, see Paper~I). Whereas most of the pulsars in double neutron star systems have their mass determined
accurately,
only a few of the $\sim\!160$ binary pulsars with WD companions have measured masses. These are listed in Table~\ref{table:spinup}.
At first sight, a handful 
of these are more massive than this previous limit of $1.44\,M_{\odot}$.
However, in most of those cases the mass determinations have large uncertainties and in all cases also include mass accreted from 
the progenitor of their WD companion, i.e. these masses are pulsar masses {\it after} the recycling phase and not neutron star 
birth masses after the SN.\\
To take this effect into account 
we have estimated upper limits for the birth mass of the neutron star
in all systems by subtracting the minimum mass needed to spin up the pulsar
using equation~(\ref{eq:deltaMfinalfit}). The results are shown in the fifth column of Table~\ref{table:spinup}. 
Since we do not know the true age of the recycled pulsars, nor
their braking index, we simply assumed in all cases that the presently observed pulsar spin period,~$P$ is related
to their original equilibrium spin period, $P_{\rm eq}$ prior to the RLDP via: 
$P=\sqrt{2} P_{\rm eq}$. For pulsars where the RLDP effect was small the initial spin period after recycling,~$P_0\simeq P_{\rm eq}$
and hence $P\simeq \sqrt{2} P_0$, which corresponds to the case where the true age is half the characteristic age
of a pulsar with braking index $n=3$ (see Section~\ref{sec:trueages} for further discussion on braking indices and true ages).

%-------------------------------------------------------------------------------
\begin{table}
\center
\caption{Measured masses of neutron stars in NS-WD systems in the Galactic disk, including PSR~J1903+0327.
         The spin periods in the second column are in ms.
         The third, fourth and fifth column state the measured mass, the minimum accreted mass needed to spin~up 
         the pulsar using equation~(\ref{eq:deltaMfinalfit}) and the derived
         upper value for its birth mass, $M_{\rm NS, max}^{\rm birth}=M_{\rm NS}-\Delta M_{\rm eq}$, respectively.
         All masses are in quoted in $M_{\odot}$. All error bars are 1-$\sigma$ values.}
\begin{tabular}{lllll}
\hline {Pulsar} & {$P_{\rm spin}$} & {$M_{\rm NS}$} & {$\Delta M_{\rm eq}$} & {$M_{\rm NS, max}^{\rm birth}$} \\ 
\hline 
\noalign{\smallskip} 
            J0437--4715$^1$    & {5.76} & $1.76\pm 0.20$ & 0.04 & $1.72\pm 0.20$\\ %Mwd=0.254 +- 0.018 (68%) Verbiest et al. (2008)
            J0621+1002$^2$     & {28.9} & $1.70^{\rm +0.10}_{\rm -0.17}$ & 0.005 & $1.70\pm 0.14$\\ % Mwd=? Nice et al. (2008)
            J0751+1807$^2$     & {3.48} & $1.26\pm 0.14$ & 0.07 & $1.19\pm 0.13$\\ % Mwd=? Nice et al. (2008)
            J1012+5307$^3$     & {5.26} & $1.64\pm 0.22$ & 0.05 & $1.59\pm 0.21$\\ %Mwd=0.16 +- 0.02 Callanan et al. (1998)
            J1141--6545$^4$    & {394}  & $1.27\pm 0.01$ & --- & $1.27\pm 0.01$\\ %Mwd=?   Bhat et al. (2008)
            J1614--2230$^5$    & {3.15} & $1.97\pm 0.04$ & 0.09 & $1.88\pm 0.04$\\ %Mwd=0.500, Demorest et al. (2010)
            J1713+0747$^6$     & {4.57} & $1.53^{\rm +0.08}_{\rm -0.06}$  & 0.05 & $1.48\pm 0.07$\\ %Mwd=? Splaver et al. (2005)
            J1738+0333$^7$     & {5.85} & $1.46^{\rm +0.06}_{\rm -0.05}$  & 0.04 & $1.42\pm 0.05$\\ %Mwd=? Antoniadis et al. (2011)
            J1802--2124$^8$    & {12.6} & $1.24\pm 0.11$ & 0.01 & $1.23\pm 0.11$\\ %Mwd=0.78 +- 0.04  Ferdnan et al. (2010)
            B1855+09$^9$       & {5.36} & $1.57^{\rm +0.12}_{\rm -0.11}$ & 0.04 & $1.53\pm 0.11$\\ %Mwd= (68%) Nice et al. (2003)
            J1903+0327$^{10}$  & {2.15} & $1.667\pm 0.021$ & 0.15 & $1.52\pm 0.02$\\ %M2=    Freire et al. (2011)
            J1909--3744$^{11}$ & {2.95} & $1.44\pm 0.02$ & 0.09 & $1.35\pm 0.02$\\ %Mwd=    Jacoby et al. (2005)
            B2303+46$^{12}$    & {1066} & $1.38^{\rm +0.06}_{\rm -0.10}$ & --- & $1.38\pm 0.08$\\ %Mwd= Thorsett & Chakrabarty (1999)
\noalign{\smallskip} 
\hline
\end{tabular}
\begin{flushleft}
References: (1) \citet{vbv+08}, (2) \citet{nsk08}, (3) \citet{cgk98}, (4) \citet{bbv08}, (5) \citet{dpr+10}, (6) \citet{sns+05},
            (7) \citet{akk+12}, (8) \citet{fsk+10}, (9) \citet{nss03}, (10) \citet{fbw+11}, (11) \citet{jhb+05}, (12) \citet{tc99}.
\end{flushleft}

\label{table:spinup}
\end{table}
%-------------------------------------------------------------------------------
It is possible that some BMSPs have $P\simeq P_0$ (especially those BMSPs with small values of $B_0$). In this case
the neutron star birth masses would be slightly larger than indicated here.
However, recall that the limits in Table~\ref{table:spinup}
are absolute upper limits obtained for {\it idealized} and efficient spin~up. Furthermore, once an accreting pulsar reaches its equilibrium
spin period it can still accrete material but the net effect may not be further spin~up (depending on the mass-transfer rate, $\dot{M}$).
Therefore, it should be emphasized that the actual SN birth mass could in many cases be substantially lower
than estimated in Table~\ref{table:spinup}.
For example, we have demonstrated in Paper~I that \psr\ could have accreted
$0.31\,M_{\odot}$, which is substantially more than shown in Table~\ref{table:spinup} where the pulsar in this case was assumed 
to be born with a spin period of $3.15\,{\rm ms}/\sqrt{2}=2.23\,{\rm ms}$, yielding $\Delta M_{\rm eq}=0.09\,M_{\odot}$.

\subsection{PSR J0621+1002}
Considering the derived SN birth masses in Table~\ref{table:spinup} we notice at first sight in particular PSR~J0437$-$4715 and PSR~J0621+1002
as further candidates for pulsars which were potentially born massive ($\sim 1.7\,M_{\odot}$), although the error bars are large.
Given the uncertainty of the accretion efficiency (i.e. the possible difference between the actual amount of mass accreted and $\Delta M_{\rm eq}$)
we see that in particular PSR~J0621+1002 could have been born massive with $M_{\rm NS, max}^{\rm birth}\sim 1.7\,M_{\odot}$.
PSR~J0621+1002 \citep{cnst96} has an orbital period of 8.3~days and a {CO}~WD companion of mass $0.67\,M_{\odot}$
which is somewhat equivalent to \psr.
However, its relatively slow spin period of 28.9~ms and large eccentricity of $2.5\times 10^{-3}$
indicate that PSR~J0621+1002 did not evolve through an {X}-ray phase with stable, and relatively long Case~A RLO (unlike the 
situation for \psr). It seems much more evident that PSR~J0621+1002 evolved through a rapid, less efficient, mass-transfer phase -- 
such as early Case~B RLO \citep{tvs00}. In this case the actual amount of accretion must have been at most a few $10^{-2}\,M_{\odot}$. Therefore,
its measured mass is very close to its birth mass and {\it if} future mass measurements of PSR~J0621+1002 
yield smaller error bars, still centered on its current mass estimate, we conclude that
this pulsar belongs to the same class of neutron stars as \psr, Vela~{X}-1 and possibly the Black-Widow pulsar which
were all born massive (see Paper~I). However, the error bar is quite large and PSR~J0621+1002 could turn out to have 
an ordinary mass of about $1.4\,M_{\odot}$ (the 2-$\sigma$ lower mass limit). 

\subsubsection{PSR J0437$-$4715}
PSR~J0437$-$4715 is also a potential candidate for a pulsar which could have been born as a neutron star with a mass of
$\sim\!1.7\,M_{\odot}$. However, this pulsar has a fairly rapid spin period (like a few other relatively high-mass 
neutron star candidates in Table~\ref{table:spinup}) which indicates a long spin-up phase. Hence, it may have
accreted significantly more than suggested by $\Delta M_{\rm eq}$. In fact, as mentioned before, in the case of \psr\ the pulsar might have 
accreted three times more than suggested by $\Delta M_{\rm eq}$, see fig.~7 in \citet{tlk11}.
The problem is that as soon as the pulsar spin reaches $P_{\rm eq}$ (which requires accumulation
of mass $\Delta M_{\rm eq}$) it may remain in near-spin equilibrium and keep accreting on and off
while spinning up/down depending on the sign of the accretion torque.
The situation is quite different for PSR~J0621+1002 which formed through a short lived {X}-ray phase excluding 
evolution at near-spin equilibrium for a substantial amount of time.
We therefore conclude that PSR~J0437$-$4715 could originally have been born in a SN with a typical mass of $\sim\!1.4\,M_{\odot}$,
even if its present mass is indeed $\sim\!1.76\,M_{\odot}$.

%%%%%%%%%%%%%%%%%%%%%%%%%%%%%%%%%%%%%%%%%%%%%%%%%%%%%%%%%%%%%%%%%%%%%%%%%%%%%%%%

\section{Summary}\label{sec:summary}
We have investigated in detail the recycling process of pulsars with respect to accretion, spin~up and the
Roche-lobe decoupling phase, and discussed the implications for their spin periods, masses and true ages.
In particular, we have discussed the concept of a spin-up line in the $P\dot{P}$--diagram and emphasize that 
such a line cannot be uniquely defined. Besides from the poorly known disk-magnetosphere physics, 
which introduces large uncertainties,  
for each individual pulsar the equilibrium spin period, $P_{\rm eq}$ also depends on its 
magnetic inclination angle, $\alpha$ as well as
its accretion history ($\dot{M}$) and its B-field strength.
Furthermore, we have applied the \citet{spi06} spin-down torque on radio pulsars which significantly
changes the location of the spin-up lines compared to using the vacuum magnetic dipole model, especially
for small values of $\alpha$. 

We have derived a simple analytical expression (equation~\ref{eq:deltaMfinalfit})
to evaluate the amount of mass needed to be accreted to spin~up a pulsar to any given equilibrium spin period, $P_{\rm eq}$. 
Our result resembles that of \citet{acrs82} and approximately yields the same values (within a factor of two)
as the expression derived by \citet{lp84}.  
Using our formula we find, for example, that BMSPs with {He}~WDs and $P_{\rm eq}\simeq2\,{\rm ms}$ must have accreted at least $0.10\,M_{\odot}$,
whereas typical BSMPs with {CO}~WDs and $P_{\rm eq}\simeq20\,{\rm ms}$ only needed to accrete $0.005\,M_{\odot}$. 

Applying equation~(\ref{eq:deltaMfinalfit}) enables us to explain
the difference in spin distributions between BMSPs with {He} and {CO}~WDs, respectively.
The BMSPs with {He}~WDs often evolved via an {X}-ray phase with stable RLO on a long timescale, allowing sufficient material to be accreted  
by the neutron star to spin it up efficiently to a short period, whereas the BMSPs with {CO}~WDs most often have slow spins and evolved from IMXBs
which had a short phase of mass transfer via early Case~B RLO or Case~C RLO -- the latter leading to a CE-evolution 
often followed by Case~BB RLO.
The only exception known so far is \psr\ since this system produced a {CO}~WD orbiting a fully recycled MSP and thus it must
have evolved via Case~A RLO of an IMXB. 

It is not possible to recycle pulsars to become MSPs via wind accretion from $1.1-2.2\,M_{\odot}$ post-CE helium stars.
However, we have demonstrated that Case~BB RLO from such helium stars can spin~up pulsars
to at least $\sim\!11~{\rm ms}$. Further studies of these systems are needed.

There is an increasing number of recycled pulsars with WD companions which seem to fall outside the
two main populations of BMSPs with He and CO~WDs, respectively. These peculiar systems possibly  
have He~WDs and always exhibit slow spin periods between 10 and 100~ms. We suggest that these systems with $P_{\rm orb}\ge 1\;{\rm day}$ may
have formed via Case~A RLO of IMXBs. We plan further studies on these binaries. 

The Roche-lobe decoupling phase (RLDP), at the terminal stage of the mass transfer \citep{tau12}, has been 
analysed and we have shown that
while the RLDP~effect is important in LMXBs -- leading to significant loss of rotational energy of the recycled pulsars
as well as characteristic ages at birth which may exceed the age of the Universe 
-- it is not
significant in IMXB systems where the duration of the RLDP is short.

In order to track the evolution of pulsars in the $P\dot{P}$--diagram we have introduced 
two types of true age isochrones -- one
which matches well with the banana shape of the observed distribution of known MSPs.
We encourage further statistical population studies to better understand the formation and evolution of radio MSPs in the
$P\dot{P}$--diagram \citep[see e.g.][]{kt10}. 
The discrepancy between true ages and characteristic spin-down ages of recycled pulsars has been discussed and we confirm 
that the latter values are completely untrustworthy as true age indicators, leaving WD cooling ages as the
only valid, although not accurate, measuring scale \citep{tau12}. 

In the combined study presented here and in Paper~I
we have investigated the recycling of \psr\ by detailed modelling of the mass exchanging {X}-ray phase of the progenitor system.
Given the rapid spin of \psr\ (3.15~ms) we argue that it is highly unlikely that it evolved through a CE, leaving Case~A RLO in an IMXB
as the only viable formation channel. 
We confirm the conclusion from Paper~I that the neutron star in \psr\ was born massive ($1.70\pm0.15\,M_{\odot}$).
We have demonstrated that \psr\ could have been spun-up at $\dot{M}=\dot{M}_{\rm Edd}$ and subsequently evolve
to its current position in the $P\dot{P}$--diagram within 2~Gyr (the estimated cooling age of its white dwarf companion).

Besides \psr, Vela~{X}-1 and possibly the Black-Widow pulsar, we have argued that also
PSR~J0621+1002 could belong to the same class of neutron stars born massive ($\ge 1.7\,M_{\odot}$). 
The formation of such massive neutron stars in supernovae is in agreement with some 
supernova explosion models \citep[e.g.][]{zwh08,ujma12}.

%%%%%%%%%%%%%%%%%%%%%%%%%%%%%%%%%%%%%%%%%%%%%%%%%%%%%%%%%%%%%%%%%%%%%%%%%%%%%%%%

\section*{Acknowledgements}
T.M.T. gratefully acknowledges financial support and hospitality at both the 
  Argelander-Insitut f\"ur Astronomie, Universit\"at Bonn
  and the Max-Planck-Institut f\"ur Radioastronomie.
%\newpage
\bibliographystyle{mn2e} 
\bibliography{tauris_refs}
%\bibliography{psrrefs,modrefs,crossrefs,journals}
%\bibliography{journals}

%%%%%%%%%%%%%%%%%%%%%%%%%%%%%%%%%%%%%%%%%%%%%%%%%%%%%%%%%%%%%%%%%%%%%%%%%%%%%%%%
\newpage
\appendix
\section{Identifying the nature of a pulsar companion star} \label{appendix}
For stellar evolutionary purposes, and for observers to judge the possibilities to detect a companion, 
it would be useful to have a tool to predict the most likely
nature of the companion star of a given binary pulsar. As discussed in Section~\ref{sec:observations}  
radio pulsars are presently found with the following companions:
\begin{itemize} 
\item Main sequence stars -- {\em MS}
\item Neutron stars -- {\em NS}
\item CO/ONeMg white dwarfs -- {\em CO}
\item He white dwarfs -- {\em He}
\item Ultra-light companions (or planets) -- {\em UL}
\end{itemize} 
Although still not detected, radio pulsars are also expected to exist in systems 
with either:
\begin{itemize} 
\item Helium stars -- {\em HS}
\item (sub)Giant stars -- {\em GS}
\item Black Holes -- {\em BH}
\end{itemize} 
As explained in Section~\ref{sec:observations} pulsar systems found in globular clusters are not useful as probes of stellar 
evolution due to perturbations in their dense environment. We shall denote the companions of such pulsars by:
\begin{itemize} 
\item Globular cluster companions -- {\em GC}
\end{itemize} 

The usual characteristic parameters measured of an observed binary pulsar are: the spin period ($P$), the period derivative ($\dot{P}$),
the orbital period ($P_{\rm orb}$) and the projected semi-major axis of the pulsar ($a_1$). From the latter two parameters one can
calculate the mass function ($f$) which is needed, but not sufficient, to estimate the mass of the companion star ($M_2$). 
Furthermore, in many cases it is also possible to estimate the eccentricity ($ecc$).
Based on these parameters we state simple conditions in Table~\ref{table:appendix} 
which can be used to identify the most likely nature of pulsar companions at the 
95~per~cent level of confidence (based on statistics from current available data). 
%-------------------------------------------------------------------------------
\begin{table*}
\center
\caption{The most likely nature of a radio pulsar companion star based on observable characteristics. 
         The selection conditions can be directly applied to
         the {\it ATNF Pulsar Catalogue} \citep{mhth05}.
         See also Section~\ref{sec:observations} for further discussions.}
\begin{tabular}{clll}
\hline {Companion type} & Conditions &  \\ 
\hline 
\noalign{\smallskip} 
  {\em MS}$\;\;$ & $M_2^{*}>0.5\;M_{\odot}$        & and & $P_{\rm orb}>50^{\rm d}$ \\ 
  {\em NS}$\;\;$ &                                 &     & $P_{\rm orb}<50^{\rm d}$     \quad  and \quad $ecc>0.05$ \\
  {\em CO}$\;\;$ & $M_2^{*}>0.335\;M_{\odot}$      & and & $P_{\rm orb}<75^{\rm d}$     \quad  and \quad $ecc<0.05$ \quad and \quad  $P>8\;{\rm ms}$ \\
  {\em He}$\;\;$ & $M_2^{*}>0.08\;M_{\odot}$       & and & \{($P_{\rm orb}<75^{\rm d}\;\;$ and $\;\;M_2^{*}<0.335\;M_{\odot}$) \quad or \quad 
                                                       ($P_{\rm orb}>75^{\rm d}\;\;$ and $\;\;M_2^{*}<0.46\;M_{\odot}$)\} \\
  {\em UL}$^{**}$ & $M_2^{*}<0.08\;M_{\odot}$ &     & \\
\noalign{\smallskip} 
\hline
\end{tabular}
\begin{flushleft}
  $^*$ The median companion mass $M_2$ is calculated for an orbital inclination
       angle $i=60^{\circ}$ and an assumed pulsar mass $M_{\rm NS}=1.35\,M_{\odot}$. \\
  $^{**}$ Many pulsars with unmeasured values of $a_1$ are also expected to host an ultra-light companion
          if $P_{\rm orb}<2\;{\rm days}$ and $P<8\;{\rm ms}$.
\end{flushleft}
\label{table:appendix}
\end{table*}
%-------------------------------------------------------------------------------
These conditions can be directly applied to the online, public available 
{\it ATNF Pulsar Catalogue}, http://www.atnf.csiro.au/research/pulsar/psrcat/ \citep{mhth05}
in order to select a subsample of pulsars with certain companion stars.

One example of the few pulsars where the
identification fails is \psr\ given its unusual combination of being fully recycled and having a CO~WD companion, as discussed in this paper. 
The identification conditions can in principle be optimised further to increase the level of confidence even more. 
However, this would not only require 
these conditions to be updated frequently with new discoveries of somewhat peculiar binary pulsars but 
also lead to undesirable complexity in the conditions. The main uncertainty factor is the (most often) unknown orbital inclination angle
of the binary system. The transition between systems with {\em UL} and {\em He} companions and, in particular, 
between {\em He} and {\em CO} companions is 
difficult to define satisfactory. Furthermore, WD+NS systems where the last formed compact object is the neutron star are very difficult
to distinguish from NS+NS systems, {\it if} based solely on data from the last formed (non-recycled) NS.
Finally, we note that the long-sought after radio pulsar in a binary with a black hole companion 
(NS+BH or BH+NS) will most likely be mildly recycled (i.e. NS+BH) due to its much longer lifetime as an observable pulsar after accretion. 
In this case it may be difficult to distinguish it from a system where the pulsar has a {\em NS} companion orbiting with a small inclination angle. 
On the other hand, if the pulsar is not recycled then such a BH+NS system could resemble a pulsar with an (undetected) {\em MS} companion.

\end{document}